\documentstyle[12pt,a4]{article}

\def\NEG#1{{\rlap/#1}}%
\makeatletter
\let\DOTSI\relax
\def\RIfM@{\relax\ifmmode}%
\def\FN@{\futurelet\next}%
\newcount\intno@
\def\iint{\DOTSI\intno@\tw@\FN@\ints@}%
\def\iiint{\DOTSI\intno@\thr@@\FN@\ints@}%
\def\iiiint{\DOTSI\intno@4 \FN@\ints@}%
\def\idotsint{\DOTSI\intno@\z@\FN@\ints@}%
\def\ints@{\findlimits@\ints@@}%
\newif\iflimtoken@
\newif\iflimits@
\def\findlimits@{\limtoken@true\ifx\next\limits\limits@true
 \else\ifx\next\nolimits\limits@false\else
 \limtoken@false\ifx\ilimits@\nolimits\limits@false\else
 \ifinner\limits@false\else\limits@true\fi\fi\fi\fi}%
\def\multint@{\int\ifnum\intno@=\z@\intdots@                                
 \else\intkern@\fi                                                          
 \ifnum\intno@>\tw@\int\intkern@\fi                                         
 \ifnum\intno@>\thr@@\int\intkern@\fi                                       
 \int}
\def\multintlimits@{\intop\ifnum\intno@=\z@\intdots@\else\intkern@\fi
 \ifnum\intno@>\tw@\intop\intkern@\fi
 \ifnum\intno@>\thr@@\intop\intkern@\fi\intop}%
\def\intic@{\mathchoice{\hskip.5em}{\hskip.4em}{\hskip.4em}{\hskip.4em}}%
\def\negintic@{\mathchoice
 {\hskip-.5em}{\hskip-.4em}{\hskip-.4em}{\hskip-.4em}}%
\def\ints@@{\iflimtoken@                                                    
 \def\ints@@@{\iflimits@\negintic@\mathop{\intic@\multintlimits@}\limits    
  \else\multint@\nolimits\fi                                                
  \eat@}
 \else                                                                      
 \def\ints@@@{\iflimits@\negintic@
  \mathop{\intic@\multintlimits@}\limits\else
  \multint@\nolimits\fi}\fi\ints@@@}%
\def\intkern@{\mathchoice{\!\!\!}{\!\!}{\!\!}{\!\!}}%
\def\plaincdots@{\mathinner{\cdotp\cdotp\cdotp}}%
\def\intdots@{\mathchoice{\plaincdots@}%
 {{\cdotp}\mkern1.5mu{\cdotp}\mkern1.5mu{\cdotp}}%
 {{\cdotp}\mkern1mu{\cdotp}\mkern1mu{\cdotp}}%
 {{\cdotp}\mkern1mu{\cdotp}\mkern1mu{\cdotp}}}%
\def\rmfam{\z@}%
\newif\iffirstchoice@
\firstchoice@true
\def\textfonti{\the\textfont\@ne}%
\def\textfontii{\the\textfont\tw@}%
\def\text{\RIfM@\expandafter\text@\else\expandafter\text@@\fi}%
\def\text@@#1{\leavevmode\hbox{#1}}%
\def\text@#1{\mathchoice
 {\hbox{\everymath{\displaystyle}\def\textfonti{\the\textfont\@ne}%
  \def\textfontii{\the\textfont\tw@}\textdef@@ T#1}}%
 {\hbox{\firstchoice@false
  \everymath{\textstyle}\def\textfonti{\the\textfont\@ne}%
  \def\textfontii{\the\textfont\tw@}\textdef@@ T#1}}%
 {\hbox{\firstchoice@false
  \everymath{\scriptstyle}\def\textfonti{\the\scriptfont\@ne}%
  \def\textfontii{\the\scriptfont\tw@}\textdef@@ S\rm#1}}%
 {\hbox{\firstchoice@false
  \everymath{\scriptscriptstyle}\def\textfonti
  {\the\scriptscriptfont\@ne}%
  \def\textfontii{\the\scriptscriptfont\tw@}\textdef@@ s\rm#1}}}%
\def\textdef@@#1{\textdef@#1\rm\textdef@#1\bf\textdef@#1\sl\textdef@#1\it}%
\def\DN@{\def\next@}%
\def\eat@#1{}%
\def\textdef@#1#2{%
 \DN@{\csname\expandafter\eat@\string#2fam\endcsname}%
 \if S#1\edef#2{\the\scriptfont\next@\relax}%
 \else\if s#1\edef#2{\the\scriptscriptfont\next@\relax}%
 \else\edef#2{\the\textfont\next@\relax}\fi\fi}%
\def\Let@{\relax\iffalse{\fi\let\\=\cr\iffalse}\fi}%
\def\vspace@{\def\vspace##1{\crcr\noalign{\vskip##1\relax}}}%
\def\multilimits@{\bgroup\vspace@\Let@
 \baselineskip\fontdimen10 \scriptfont\tw@
 \advance\baselineskip\fontdimen12 \scriptfont\tw@
 \lineskip\thr@@\fontdimen8 \scriptfont\thr@@
 \lineskiplimit\lineskip
 \vbox\bgroup\ialign\bgroup\hfil$\m@th\scriptstyle{##}$\hfil\crcr}%
\def\Sb{_\multilimits@}%
\def\endSb{\crcr\egroup\egroup\egroup}%
\def\Sp{^\multilimits@}%

\newdimen\ex@
\def\rightarrowfill@#1{$#1\m@th\mathord-\mkern-6mu\cleaders
 \hbox{$#1\mkern-2mu\mathord-\mkern-2mu$}\hfill
 \mkern-6mu\mathord\rightarrow$}%
\def\leftarrowfill@#1{$#1\m@th\mathord\leftarrow\mkern-6mu\cleaders
 \hbox{$#1\mkern-2mu\mathord-\mkern-2mu$}\hfill\mkern-6mu\mathord-$}%
\def\leftrightarrowfill@#1{$#1\m@th\mathord\leftarrow\mkern-6mu\cleaders
 \hbox{$#1\mkern-2mu\mathord-\mkern-2mu$}\hfill
 \mkern-6mu\mathord\rightarrow$}%
\def\overrightarrow{\mathpalette\overrightarrow@}%
\def\overrightarrow@#1#2{\vbox{\ialign{##\crcr\rightarrowfill@#1\crcr
 \noalign{\kern-\ex@\nointerlineskip}$\m@th\hfil#1#2\hfil$\crcr}}}%

\def\overleftarrow{\mathpalette\overleftarrow@}%
\def\overleftarrow@#1#2{\vbox{\ialign{##\crcr\leftarrowfill@#1\crcr
 \noalign{\kern-\ex@\nointerlineskip}$\m@th\hfil#1#2\hfil$\crcr}}}%
\def\overleftrightarrow{\mathpalette\overleftrightarrow@}%
\def\overleftrightarrow@#1#2{\vbox{\ialign{##\crcr\leftrightarrowfill@#1\crcr
 \noalign{\kern-\ex@\nointerlineskip}$\m@th\hfil#1#2\hfil$\crcr}}}%
\def\underrightarrow{\mathpalette\underrightarrow@}%
\def\underrightarrow@#1#2{\vtop{\ialign{##\crcr$\m@th\hfil#1#2\hfil$\crcr
 \noalign{\nointerlineskip}\rightarrowfill@#1\crcr}}}%

\def\underleftarrow{\mathpalette\underleftarrow@}%
\def\underleftarrow@#1#2{\vtop{\ialign{##\crcr$\m@th\hfil#1#2\hfil$\crcr
 \noalign{\nointerlineskip}\leftarrowfill@#1\crcr}}}%
\def\underleftrightarrow{\mathpalette\underleftrightarrow@}%
\def\underleftrightarrow@#1#2{\vtop{\ialign{##\crcr$\m@th\hfil#1#2\hfil$\crcr
 \noalign{\nointerlineskip}\leftrightarrowfill@#1\crcr}}}%
\def\binom#1#2{{#1 \choose #2}}%
\long\def\QQQ#1#2{\long\expandafter\def\csname#1\endcsname{#2}}%
\def\QTP#1{}%
\long\def\QQA#1#2{}%
\def\QTR#1#2{{\csname#1\endcsname #2}}
\long\def\TeXButton#1#2{#2}%
\def\EXPAND#1[#2]#3{}%
\def\NOEXPAND#1[#2]#3{}%
\def\LaTeXparent#1{}%
\def\QTagDef#1#2#3{}%
\def\QQfnmark#1{\footnotemark}

\@ifundefined{INDEX}{\def\INDEX#1#2{}{}}{}%
\@ifundefined{SUBINDEX}{\def\SUBINDEX#1#2#3{}{}{}}{}%
\def\initial#1{\bigbreak{\raggedright\large\bf #1}\kern 2\p@\penalty3000}%
\@ifundefined{abstract}{%
 \def\abstract{%
  \if@twocolumn
   \section*{Abstract (Not appropriate in this style!)}%
   \else \small
   \begin{center}{\bf Abstract\vspace{-.5em}\vspace{\z@}}\end{center}%
   \quotation
   \fi
  }%
 }{%
 }%
\@ifundefined{endabstract}{\def\endabstract
  {\if@twocolumn\else\endquotation\fi}}{}%
\@ifundefined{maketitle}{\def\maketitle#1{}}{}%
\@ifundefined{affiliation}{\def\affiliation#1{}}{}%
\@ifundefined{proof}{}{}%
\@ifundefined{endproof}{}{}%
\@ifundefined{newfield}{\def\newfield#1#2{}}{}%
\@ifundefined{chapter}{\def\chapter#1{\par(Chapter head:)#1\par }%
 \newcount\c@chapter}{}%
\@ifundefined{part}{\def\part#1{\par(Part head:)#1\par }}{}%
\@ifundefined{section}{\def\section#1{\par(Section head:)#1\par }}{}%
\@ifundefined{subsection}{\def\subsection#1%
 {\par(Subsection head:)#1\par }}{}%
\@ifundefined{subsubsection}{\def\subsubsection#1%
 {\par(Subsubsection head:)#1\par }}{}%
\@ifundefined{paragraph}{\def\paragraph#1%
 {\par(Subsubsubsection head:)#1\par }}{}%
\@ifundefined{subparagraph}{\def\subparagraph#1%
 {\par(Subsubsubsubsection head:)#1\par }}{}%
\@ifundefined{therefore}{}{}%
\@ifundefined{backepsilon}{}{}%
\@ifundefined{yen}{}{}%
\@ifundefined{registered}{\def\registered{\relax\ifmmode{}\r@gistered
                                                \else$\m@th\r@gistered$\fi}%
 \def\r@gistered{^{\ooalign
  {\hfil\raise.07ex\hbox{$\scriptstyle\rm\text{R}$}\hfil\crcr
  \mathhexbox20D}}}}{}%
\@ifundefined{Eth}{}{}%
\@ifundefined{eth}{}{}%
\@ifundefined{Thorn}{}{}%
\@ifundefined{thorn}{}{}%
\@ifundefined{degree}{}{}%
\def\BibTeX{{\rm B\kern-.05em{\sc i\kern-.025em b}\kern-.08em
    T\kern-.1667em\lower.7ex\hbox{E}\kern-.125emX}}%
\newdimen\theight
\def\Column{%
 \vadjust{\setbox\z@=\hbox{\scriptsize\quad\quad tcol}%
  \theight=\ht\z@\advance\theight by \dp\z@\advance\theight by \lineskip
  \kern -\theight \vbox to \theight{%
   \rightline{\rlap{\box\z@}}%
   \vss
   }%
  }%
 }%
\def\qed{%
 \ifhmode\unskip\nobreak\fi\ifmmode\ifinner\else\hskip5\p@\fi\fi
 \hbox{\hskip5\p@\vrule width4\p@ height6\p@ depth1.5\p@\hskip\p@}%
 }%
\def\miss{\hbox{\vrule height2\p@ width 2\p@ depth\z@}}%
\def\tcol#1{{\baselineskip=6\p@ \vcenter{#1}} \Column}  %
\makeatother

\QQQ{Language}{
American English
}

\begin{document}

\setcounter{page} {0}
\author{Martin Moj\v zi\v s\bigskip\bigskip \\
Department of Theoretical Physics, Comenius University\\
Mlynska dolina, SK-84215 Bratislava, Slovakia\\
{\sl email: mojzis@fmph.uniba.sk}\bigskip}
\title{Elastic $\pi $N Scattering to ${\cal O}(p^3)$\\
in Heavy Baryon Chiral Perturbation Theory\thanks{%
Work supported in part by VEGA (Slovakia) grant No. 1/1323/96 and by FWF
(Austria), Project No. P09505-PHY.}\bigskip\bigskip \bigskip\ }
\date{ }
\maketitle

\begin{abstract}
The elastic $\pi $N scattering amplitude in the isospin limit is calculated
in the framework of heavy baryon chiral perturbation theory, up to the third
order. Threshold parameters like scattering lengths, volumes, effective
ranges, etc. are compared with data. All relevant low energy constants are
fixed from the available pion-nucleon data. A clear improvement in the
description of data is observed, when going from the first two orders in the
chiral expansion to the third one. The importance of even higher orders is
suggested by the results.
\end{abstract}

\TeXButton{TeX field}{\thispagestyle{empty}}\newpage

\section{Introduction}

Constraints of chiral symmetry on pion-nucleon interactions were
investigated in the sixties in terms of current algebra, and the prediction
of the S-wave $\pi $N scattering lengths \cite{W66} was one of the most
important results within this approach. In the eighties, the systematic
method --- called chiral perturbation theory (CHPT) --- of calculating
corrections to the current algebra results was invented \cite{GL84}, and it
was applied to $\pi $N scattering up to order ${\cal O}(p^3)$ \cite{GSS88}.

However, CHPT with nucleons is not as systematic as CHPT for light mesons,
since the nucleon mass spoils one of the main virtues of CHPT --- one-to-one
correspondence between loop expansion and expansion in external momenta.
This correspondence is valid for massless particles \cite{W79}, so it works
for the Goldstone bosons of the spontaneously broken chiral symmetry of QCD
(pions, kaons and $\eta $), which are massless in the chiral limit of
massless $u,$ $d$ and $s$ quarks. Nonzero masses of light quarks, leading to
nonzero masses of Goldstone bosons, are treated as a perturbation. The
simultaneous expansion in external momenta and light quark masses is called
chiral expansion. Nucleons, on the other hand, are massive even in the
chiral limit and their masses cannot be treated as small perturbations. As a
consequence, there is no more direct correspondence between loop and chiral
expansions, and diagrams with arbitrary number of loops can contribute to a
given chiral order.

This drawback was eliminated in the clever reformulation of CHPT with
baryons \cite{JM91} --- called heavy baryon chiral perturbation theory
(HBCHPT) --- which, so to say, shifts the nucleon mass from the propagator
to vertices of an effective Lagrangian and thus restores the loop--chiral
correspondence. HBCHPT was applied to different processes up to order ${\cal %
O}(p^3)$ \cite{BKM95} (and even ${\cal O}(p^4)$ \cite{BKM95b}\cite{BM96}),
among others to the forward threshold $\pi $N scattering, from where the
corrections to the S-wave $\pi $N scattering lengths up to order ${\cal O}%
(p^3)$ were calculated \cite{BKM93}. Recently also higher $\pi $N partial
waves were calculated up to this order \cite{BKM96}.

In these works, some of the low energy constants (LECs) of the ${\cal O}%
(p^3) $ $\pi $N Lagrangian were not determined from the fit to the $\pi $N
data, instead they were estimated from the principle of resonance
saturation, which had been shown to work very well in the mesonic sector 
\cite{EGPdR89}, but in the baryonic sector it is just a working hypothesis.

The purpose of this paper is to calculate the full $\pi $N scattering
amplitude in the isospin limit in HBCHPT to order ${\cal O}(p^3)$ and to fix
the LECs contributing to this amplitude from the available experimental
information. The paper is organized as follows. In section 2 we summarize
the previous results, needed for the calculation of the amplitude, in
section 3 the amplitude is calculated and in section 4 it is confronted with
data. Conclusions are summarized in section 5, and some technical points
and/or lengthy formulae are left to appendices.

\section{HBCHPT of the pion-nucleon system}

\subsection{Effective Lagrangian}

Our calculation of the elastic $\pi $N scattering is based on the low-energy
expansion of the $\pi $N Lagrangian in HBCHPT 
\begin{equation}
\label{LMB}{\cal L}_{\pi \pi }+\widehat{{\cal L}}_{\pi N}={\cal L}_{\pi \pi
}^{(2)}+{\cal L}_{\pi \pi }^{(4)}+\ldots +\widehat{{\cal L}}_{\pi N}^{(1)}+%
\widehat{{\cal L}}_{\pi N}^{(2)}+\widehat{{\cal L}}_{\pi N}^{(3)}+\ldots
\quad .
\end{equation}

The pion isotriplet field $\stackrel{\rightarrow }{\Phi }$ is represented by
the field $u(x)$ or $U(x),$ where $U=u^2.$ In the so-called sigma
parametrization 
\begin{equation}
\label{u}U(x)=\sqrt{1-\frac{\overrightarrow{\Phi }^2(x)}{F^2}}+i\frac{%
\stackrel{\rightarrow }{\tau }\cdot \stackrel{\rightarrow }{\Phi }(x)}F
\end{equation}
where $\stackrel{\rightarrow }{\tau }$ are Pauli matrices and $F$ is a LEC
of the meson Lagrangian ${\cal L}_{\pi \pi }^{(2)}$ (pion decay constant in
the chiral limit). The nucleon isodoublet field $\Psi =\binom pn$ is
decomposed using a time-like unit 4-vector $v$%
\begin{equation}
\label{Nv}
\begin{array}{rcl}
N_v(x) & = & \exp [imv\cdot x]P_v^{+}\Psi (x) \\
H_v(x) & = & \exp [imv\cdot x]P_v^{-}\Psi (x) \\
P_v^{\pm } & = & \frac 12(1\pm \!\NEG v)~,\qquad v^2=1
\end{array}
\end{equation}
(where $m$ is the nucleon mass in the chiral limit) and the ''heavy
component'' $H_v$ is integrated out \cite{MRR92}.

The Lagrangian (\ref{LMB}) was constructed from the following building
blocks:
\begin{equation}
\label{BB}
\begin{array}{rcl}
u_\mu & = & i\{u^{\dagger }(\partial _\mu -ir_\mu )u-u(\partial _\mu -i\ell
_\mu )u^{\dagger }\} \\
\Gamma _\mu & = & \frac 12\{u^{\dagger }(\partial _\mu -ir_\mu )u+u(\partial
_\mu -i\ell _\mu )u^{\dagger }\} \\
\chi _{\pm } & = & 2B\{u^{\dagger }(s+ip)u^{\dagger }\pm u(s+ip)^{\dagger
}u\} \\
\nabla _\mu & = & \partial _\mu +\Gamma _\mu -iv_\mu ^{(s)} \\
& \vdots &
\end{array}
\end{equation}
where $B$ is a LEC of the meson Lagrangian ${\cal L}_{\pi \pi }^{(2)}$ and $%
s,$ $p,$ $\ell ^\mu ,$ $r^\mu ,$ $v_\mu ^{(s)}$ are external fields (scalar
and pseudoscalar, left-handed and right-handed vector isotriplet and vector
isosinglet, respectively). For the elastic $\pi $N scattering one can set
these fields to zero (dots correspond to terms vanishing in such a case)
with the only exception of the scalar field. Via this field the nonzero
quark masses are taken into account, and in the isospin limit\footnote{%
\ From now on we shall work within the isospin limit, without stating it
always explicitly.} $m_u=m_d$ one has \cite{GL84}
\begin{equation}
\label{s}2Bs=M^2
\end{equation}
where $M$ is the bare pion mass.

\begin{table} \centering
  \begin{tabular}{|c|c|c|} \hline
     {\bf $i$}  &  {\bf $P_i$}  &  {\bf $\gamma _i$} \\ \hline
     $1$       & $4\left\langle u\cdot u\right\rangle ^2$       & $1/3$  \\ \hline
     $2$       & $4\left\langle u^\mu u^\nu \right\rangle \left\langle u_\mu u_\nu \right\rangle $       & $2/3$  \\ \hline
     $3$       &$\langle \chi _{+}\rangle ^2$      &  $-1/2$  \\ \hline
     $4$       & $2\left\langle \chi _{+}\right\rangle \left\langle u\cdot u\right\rangle
+2\left\langle \chi _{-}^2\right\rangle -\left\langle \chi _{-}\right\rangle ^2$ & $2$ \\ \hline
  \end{tabular}
  \caption{The field monomials $P_i$ in $\QTR{cal}{L}_{\pi \pi }^{(4)}$ contributing
to the elastic $\pi$N scattering. \label{Tab P}}
\end{table}

\begin{table} \centering
  \begin{tabular}{|c|c|c|} \hline
     {\bf $i$}  &  {\bf $O_i$}  &  {\bf $\beta _i$} \\ \hline
    $1$  &  $i[u_\mu ,[v\cdot \nabla, u^\mu ]]$      &  $-\dot g_A^4/6$  \\ \hline
    $2$  &  $i[u_\mu ,[\nabla ^\mu ,v\cdot u]]$       &  $-(1+5\dot g_A^2)/12$  \\ \hline
    $3$  &  $i[v\cdot u,[v\cdot \nabla ,v\cdot u]]$      &  $(3+\dot g_A^4)/6$ \\ \hline
    $4$  &  $i\langle u_\mu v\cdot u\rangle \nabla ^\mu +\text{h.c.}$      &  $0$ \\ \hline
    $6$  &  $[\chi _{-},v\cdot u]$       &  $(1+5\dot g_A^2)/24$  \\ \hline
    $15$ &  $\varepsilon ^{\mu \nu \rho \sigma }v_\rho S_\sigma \langle [v\cdot \nabla ,u_\mu ]u_\nu \rangle $       & $\dot g_A^4/3$  \\ \hline
    $16$ &  $\varepsilon ^{\mu \nu \rho \sigma }v_\rho S_\sigma \langle u_\mu [\nabla _\nu ,v\cdot u]\rangle $      & $0$  \\ \hline
    $17$ &  $S\cdot u\langle \chi _{+}\rangle $       & $\dot g_A/2+\dot g_A^3$ \\ \hline
    $19$  & $iS^\mu [\nabla _\mu ,\chi _{-}]$       &  $0$  \\ \hline
  \end{tabular}
  \caption{The field monomials $O_i$ in $\widehat{\QTR{cal}{L}}_{\pi N}^{(3)}$
contributing to the elastic $\pi$N scattering.\label{Tab O}}
\end{table}

Particular terms in (\ref{LMB}) are \cite{GSS88}\cite{EM96}
\begin{equation}
\label{L123}
\begin{array}{rcl}
{\cal L}_{\pi \pi }^{(2)} & = & \frac{F^2}4\left\langle u\cdot u+\chi
_{+}\right\rangle \\  &  &  \\
{\cal L}_{\pi \pi }^{(4)} & = & \frac 1{16}\sum_i l_i P_i \\
&  &  \\
\widehat{{\cal L}}_{\pi N}^{(1)} & = & \bar N_v\left( iv\cdot \nabla +\dot
g_AS\cdot u\right) N_v \\
&  &  \\
\widehat{{\cal L}}_{\pi N}^{(2)} & = & \bar N_v\frac 1m\left( -\frac
12(\nabla \cdot \nabla +i\dot g_A\{S\cdot \nabla ,v\cdot u\})+a_1\langle
u\cdot u\rangle \right. \\
& + & \left. a_2\langle (v\cdot u)^2\rangle +a_3\langle \chi _{+}\rangle
+a_5i\varepsilon ^{\mu \nu \rho \sigma }v_\rho S_\sigma u_\mu u_\nu +\ldots
\right) N_v \\
&  &  \\
\widehat{{\cal L}}_{\pi N}^{(3)} & = & \bar N_v\left(
\frac{\dot g_A}{8m^2}[\nabla _\mu ,[\nabla ^\mu ,S\cdot u]]+\frac
1{2m^2}\left[ \left( a_5-\frac{1-3\dot g_A^2}8\right) i\varepsilon ^{\mu \nu
\rho \sigma }u_\mu u_\nu S_\sigma i\nabla _\rho \right. \right. \\  & + &
\left.
\frac{\dot g_A}2S\cdot \nabla \;u\cdot \nabla +\frac{\dot g_A^2}8\{v\cdot
u,u_\mu \}i\varepsilon ^{\mu \nu \rho \sigma }v_\rho S_\sigma i\nabla _\nu +%
\text{h.c.}\right] \\  & + & \left. \frac 1{(4\pi F)^2}\sum_ib_iO_i+\ldots
\right) N_v
\end{array}
\end{equation}
where $S_\mu =\frac i2\gamma _5\sigma _{\mu \nu }v^\nu $ is the spin matrix, 
$\dot g_A$ is the neutron decay constant in the chiral limit, $\left\langle
.\right\rangle $ denotes trace, and dots correspond to terms not
contributing to the elastic $\pi $N scattering in the isospin limit. The
field monomials $P_i$ and $O_i$ together with their $\gamma _i$ and $\beta
_i $ (defined below) are collected in Tables \ref{Tab P} and \ref{Tab O}.

Some of the LECs are divergent and are decomposed in the standard way (using
dimensional regularization) as
\begin{equation}
\label{lr br}
\begin{array}{rcl}
l_i & = & l_i^r(\mu )+\gamma _iL(\mu ) \\ 
b_i & = & b_i^r(\mu )+\left( 4\pi \right) ^2\beta _iL(\mu )
\end{array}
\end{equation}
where 
\begin{equation}
\label{L}
\begin{array}{rcl}
L(\mu ) & = & \frac{\mu ^{D-4}}{(4\pi )^2}\left\{ \frac 1{D-4}-\frac 12[\ln
4\pi +1+\Gamma ^{\prime }(1)]\right\} \quad .
\end{array}
\end{equation}
The scale dependence of $l_i^r(\mu )$ and $b_i^r(\mu )$ is given by
\begin{equation}
\label{RG}
\begin{array}{rcl}
l_i^r(\mu ) & = & l_i^r(\mu _0)+
\frac{\gamma _i}{(4\pi )^2}\ln \frac{\mu _0}\mu  \\ b_i^r(\mu ) & = & 
b_i^r(\mu _0)+\beta _i\ln \frac{\mu _0}\mu \quad .
\end{array}
\end{equation}
In what follows we shall often use $\mu _0=M_\pi $ and express results in
terms of $\bar l_i$ and $\widetilde{b}_i$ defined by
\begin{equation}
\label{bar}
\begin{array}{rcl}
l_i^r(\mu ) & = & \frac{\gamma _i}{\left( 4\pi \right) ^2}\left( \frac{\bar
l_i}2+\ln \frac{M_\pi }\mu \right)  \\ b_i^r(\mu ) & = & \widetilde{b}%
_i+\beta _i\ln \frac{M_\pi }\mu \quad .
\end{array}
\end{equation}

Finally let us note that the nucleon field $N_v$ used in (\ref{L123}) is not
exactly the one defined in (\ref{Nv}), but is rather obtained from (\ref{Nv}%
) by so-called EOM (equation of motion) transformations \cite{EM96}. However
these EOM transformations have no effect on S-matrix elements.

\subsection{Renormalization}

To express the bare constants in ${\cal L}_{\pi \pi }^{(2)}$ and $\widehat{%
{\cal L}}_{\pi N}^{(1)}$ in terms of measurable quantities, one has to
calculate one loop corrections (using these lowest order Lagrangians) to the
pion and nucleon propagators, as well as to the coupling of the external
axial field $a^\mu =(r^\mu -\ell ^\mu )/2$ to a pion and to two nucleons.
Tree contributions from the higher-order Lagrangians should be included as
well. The results for pions are
\begin{equation}
\label{M Z F}
\begin{array}{lcl}
M^2 & = & M_\pi ^2\left( 1+
\frac{M_\pi ^2}{32\pi ^2F_\pi ^2}\bar l_3\right)  \\ Z_\pi  & = & 1-
\frac{M_\pi ^2}{F_\pi ^2}\left( 2l_4^r+6L(\mu )+\frac 1{8\pi ^2}\ln \frac{%
M_\pi }\mu \right)  \\ F & = & F_\pi \left( 1-\frac{M_\pi ^2}{16\pi ^2F_\pi
^2}\bar l_4\right) \;
\end{array}
\end{equation}
and for a nucleon with momentum $mv+k$%
\begin{equation}
\label{m ZN gA}
\begin{array}{rcl}
m & = & m_N-\delta m \\
Z_N(k) & = & 1-
\frac{\delta m}{m_N}+\frac{(k-\delta m.v)^2}{4m_N^2}-\frac{3g_A^2M_\pi ^2}{%
4F_\pi ^2}\left( 6L(\mu )+\frac 1{8\pi ^2}+\frac 3{8\pi ^2}\ln \frac{M_\pi }%
\mu \right)  \\ \dot g_A & = & g_A\left( 1+\frac{\delta m}{m_N}\right) +%
\frac{M_\pi ^2}{16\pi ^2F_\pi ^2}\left( g_A^3-4\widetilde{b}_{17}\right)
\end{array}
\end{equation}
where
\begin{equation}
\label{delta m}\delta m=-\frac{4M_\pi ^2}{m_N}a_3-\frac{3g_A^2M_\pi ^3}{%
32\pi F_\pi ^2}\quad .
\end{equation}

Let us comment on the results for nucleons, since they are different from
what was used in most of previous calculations (see e.g. \cite{BKKM92}).
The first difference is the
term $\delta m/m_N$ in $Z_N$-factor and $\dot g_A.$ The reason of this
difference is, as pointed out in \cite{FLMS97}, that the Lagrangian commonly
used in previous calculations contains some terms, which were transformed
away in (\ref{L123}) by EOM transformations \cite{EM96}. The second
difference is the term $(k-\delta m.v)^2/4m_N$ in $Z_N$-factor, which
accounts for the contribution of heavy component of the nucleon source. This
issue was not discussed in the literature on HBCHPT yet, so we shall briefly
comment on it. More systematic analysis is to be found in \cite{EM97}.

In the path integral derivation of HBCHPT Lagrangian \cite{MRR92}, one
starts from the generating functional of Green functions with at most two
nucleons
\begin{equation}
\label{Z eta}Z[\eta ,\overline{\eta }]=\int \left[ {\cal D}\Psi {\cal D}%
\overline{\QTR{(Normal)}{\Psi }}\QTR{(Normal)}{{\cal D}u}\right]
\QTR{(Normal)}{\;}\exp i\left\{ S_{\pi \pi }+S_{\pi N}+\int d^4x\;\left(
\overline{\eta }\Psi +\overline{\Psi }\eta \right) \right\} \quad .
\end{equation}
After decomposition of the nucleon field to heavy and light component (\ref
{Nv}) and similar decomposition of the nucleon source 
\begin{equation}
\label{rho}
\begin{array}{rcl}
\rho _v(x) & = & \exp [imv\cdot x]P_v^{+}\eta (x) \\
R_v(x) & = & \exp [imv\cdot x]P_v^{-}\eta (x)
\end{array}
\end{equation}
one writes 
\begin{equation}
\label{S piN}S_{\pi N}=\int d^4x\;\left\{ \overline{N}_vAN_v+\overline{H}%
_vBN_v+\overline{N}_vB^{\prime }H_v-\overline{H}_vCH_v\right\} 
\end{equation}
where $B^{\prime }=\gamma ^0B^{\dagger }\gamma ^0.$ After Gaussian
integration over $\left[ {\cal D}H_v{\cal D}\overline{H}_v\right] $ one gets 
\begin{equation}
\label{Z rho}
\begin{array}{c}
Z[\rho _v,R_{v,}
\overline{\rho }_v,\overline{R}_v]=\int \left[ {\cal D}N{\cal D}\overline{N}%
\QTR{(Normal)}{{\cal D}u}\right] \QTR{(Normal)}{\;\Delta }_H\;\exp i\left\{
S_{\pi \pi }+\widehat{S}_{\pi N}+\text{\QTR{(Normal)}{sources}}\right\}  \\  
\\ 
\widehat{S}_{\pi N}=\int d^4x\;\widehat{{\cal L}}_{\pi N}=\int d^4x\;%
\overline{N}_v\left( A+B^{\prime }C^{-1}B\right) N_v \\  \\
\text{\QTR{(Normal)}{sources}}=\int d^4x\;\left[ \left( \overline{\rho }_v+%
\overline{R}_vC^{-1}B\right) N_v+\overline{N}_v\left( \rho _v+B^{\prime
}C^{-1}R_v\right) +\overline{R}_vC^{-1}R_v\right]
\end{array}
\end{equation}
where the determinant $\QTR{(Normal)}{\Delta }_H$ turns out to be just a
constant.

In S-matrix elements, the source $\eta $ is, so to say, replaced by a Dirac
bispinor $u(mv+k,\sigma)$ (for a nucleon with momentum $mv+k$ and spin $\sigma$).
The corresponding light and heavy components lead to
\begin{equation}
\label{uH}
\begin{array}{ccccc}
\rho _v & \to & P_v^{+}u(mv+k,\sigma ) & = & \left( 1+
\frac{\delta m}{2m_N} \NEG v-\frac 1{2m_N}\NEG k\right) u(mv+k,\sigma
) \\ R_v & \to & P_v^{-}u(mv+k,\sigma ) & = & \left( -\frac{\delta m}{2m_N}\NEG
v+\frac 1{2m_N}\NEG k\right) u(mv+k,\sigma )\quad .
\end{array}
\end{equation}
Clearly, $\rho _v$ leads to a quantity of the zeroth, and $R_v$ of the first
chiral order. The lowest chiral order in the products $C^{-1}B$ and $%
B^{\prime }C^{-1}$ is also the first one. So the heavy source terms in (\ref
{Z rho}) contribute two chiral orders higher than the light source ones,
and they start to contribute at the third chiral order.

To incorporate heavy sources into calculations of various amplitudes does
not require any extra effort. For every Feynman diagram with light sources,
one just adds diagrams with one or both light sources replaced by the heavy
ones - the only differences in corresponding amplitudes are different
factors for nucleon external legs. As a matter of fact, the whole effect of
heavy sources is finally reduced to the momentum dependent, but otherwise
very simple term in $Z_N$-factor. Using $Z_N$ given by (\ref{m ZN gA}), one
can calculate with just light sources and forget completely about heavy
ones, since all their effect is taken into account in this $Z_N$ \cite{EM97}.

Heavy sources were not considered explicitly in most of previous
calculations. The reason is that in the third order they only appear as the $%
Z_N$ corrections to the first order, and the first order used to be treated
in a special way --- namely one used the original ''relativistic''
Lagrangian ${\cal L}_{\pi N}^{(1)}$ and then expanded the result in $\mu
=M_\pi /m_N.$ Advantage of this approach is that it gives automatically not
only the lowest order amplitude, but also higher order corrections to it,
including the discussed part of the $Z_N$ correction. Disadvantage, from our
point of view, is that part of these higher order corrections is included
in the definition of LECs of higher orders. So to avoid double counting in
''relativistic 1-st order'' approach, one has to know precisely what portion
of LECs comes from the lowest order Lagrangian. This is well known for the
second order LECs, but not for the third order ones, and if one does not
want to work this out, it is better to perform the whole calculation
strictly within HBCHPT.

However, we can use ''relativistic 1-st order'' approach to check our
prescription for the treatment of heavy sources. Let us take e.g. Born term
for the elastic $\pi $N-scattering (i.e. first order term with the nucleon
propagator). ''Relativistic 1-st order'' calculation gives for the invariant
amplitude $A^{+}$ defined below in (\ref{TpN}) 
\begin{equation}
\label{A rel}A_{rel}^{+}=\frac{g_A^2m_N}{F_\pi ^2}\quad . 
\end{equation}
On the other hand, HBCHPT calculation of the same term in tree approximation
up to the third order, ignoring heavy sources and expanding in $\mu ,$ gives
(for a kinematics defined by (\ref{v}), (\ref{kin}))
\begin{equation}
\label{A HBCHPT}A^{+}=\frac{g_A^2}{F_\pi ^2}\left( m_N-\frac t{8m_N}\right) +%
{\cal \ldots }
\end{equation}
and heavy source contribution, which turns out to give just an overall
multiplicative factor $\left( 1+t/8m_N^2\right) ,$ brings these two results
in agreement. For the invariant amplitude $B^{+}$ one has the same
situation, since also here one gets 
\begin{equation}
\label{B Brel}B^{+}=B_{rel}^{+}\left( 1-\frac t{8m_N^2}\right) +{\cal \ldots
\quad .} 
\end{equation}
For the invariant amplitudes $A^{-}$ and $B^{-},$ ''relativistic 1-st
order'' and HBCHPT results agree to order ${\cal O}(\mu ^2)$ even without
heavy sources contribution, and this agreement is not spoiled by the heavy
sources correction, because it starts to contribute only at order ${\cal O}%
(\mu ^3).$

\pagebreak 

\section{Elastic $\pi $N scattering}

\subsection{Kinematics}

For the elastic $\pi $N scattering we use the following kinematics: ingoing
pion with momentum $q$ and isospin $a,$ outgoing pion with momentum $%
q^{\prime }$ and isospin $b,$ ingoing nucleon with momentum $mv+p,$ and
outgoing nucleon with momentum $mv+p^{\prime }$ 
\begin{equation}
\label{scatt}\pi ^a(q)+N(mv+p)\rightarrow \pi ^b(q^{\prime })+N(mv+p^{\prime
})\quad . 
\end{equation}
The natural (and advantageous) choice for the 4-vector $v$ is the 4-velocity
of the ingoing (on-shell) nucleon
\begin{equation}
\label{v}m_Nv=mv+p\quad .
\end{equation}
In such a case one has $p=\delta m.v,$ where $\delta m$ is the sum of small
quantities of the second and third orders, and so $p$ itself is also a small
quantity of the second order.

The scattering amplitude $T$ is given in the form 
\begin{equation}
\label{T}
\begin{array}{c}
T=T^{+}\delta ^{ab}-T^{-}i\varepsilon ^{abc}\tau ^c \\  
\\ 
T^{\pm }=\overline{u}_N(p^{\prime },\sigma ^{\prime })\left( \alpha ^{\pm
}+\beta ^{\pm }i\varepsilon ^{\mu \nu \rho \omega }q_\mu q_\nu ^{\prime
}v_\rho S_\omega \right) u_N(p,\sigma )\quad .
\end{array}
\end{equation}
where
\begin{equation}
\label{uN}
\begin{array}{lclcl}
u_N(p,\sigma ) & = & P_v^{+}u(mv+p,\sigma )\; & = & u(mv+p,\sigma )\; \\
u_N(p^{\prime },\sigma ^{\prime }) & = & P_v^{+}u(mv+p^{\prime },\sigma
^{\prime })\; & = & \left[ 1+\frac 1{2m_N}(\!\!\NEG p-\NEG p^{^{\prime
}})\right] u(mv+p^{\prime },\sigma ^{\prime })\quad .
\end{array}
\end{equation}

The kinematics enables a significant simplification in the highest order
(third one in our case) of the chiral expansion of the scattering amplitude
. It is based on the fact that for the on-shell outgoing nucleon, $v\cdot
p^{\prime }$ is of the chiral order $2$ (although $p^{\prime }=p+q-q^{\prime
}$ itself is of the chiral order $1$). 
\begin{equation}
\label{vp'}m_N^2=(mv+p^{\prime })^2\qquad \Rightarrow \qquad v\cdot
p^{\prime }=\frac{m_N^2-m^2}{2m}-\frac{p^{\prime 2}}{2m}={\cal O}(p^2)\quad .
\end{equation}
So one can neglect $v\cdot p^{\prime }$ (as well as $v\cdot p$) in the
highest-order amplitude. Or in other words, one can replace every $v\cdot
q^{\prime }$ by $v\cdot q$ in the highest-order. The error thus introduced
is of even higher chiral order, beyond the level of precision of the
calculation at hand.

For sake of brevity, the following notation is used $w\equiv v\cdot q,$\ $%
w^{\prime }\equiv v\cdot q^{\prime }$ and if needed, one can express the
amplitude in the usual Mandelstam variables $s=(m_Nv+q)^2,$ $t=(q-q^{\prime
})^2$ and $u=(m_Nv-q^{\prime })^2$ using 
\begin{equation}
\label{kin}
\begin{array}{lcl}
q\cdot q^{\prime } & = & M_\pi ^2-\frac t2 \\ 
v\cdot q & = & \frac 1{2m_N}\left( s-m_N^2-M_\pi ^2\right)  \\ 
v\cdot q^{\prime } & = & \frac 1{2m_N}\left( s-m_N^2-M_\pi ^2+t\right) \quad
.
\end{array}
\end{equation}
\newpage\ 

\subsection{Scattering amplitude}

We shall present the amplitude order by order in the chiral expansion 
\begin{equation}
\label{orders}T=T_{(1)}+T_{(2)}+T_{(3)}\quad .
\end{equation}

$T_{(1)}$ is given by the tree graphs (Fig. 1) containing vertices defined by
$\widehat{{\cal L}}_{\pi N}^{(1)}$ (labelled as "1-st" in Figures 1--3),
with the bare coupling constants
and masses ($F,$ $\dot g_A,$ $M$ and $m$) replaced by their physical values (%
$F_\pi ,$ $g_A,$ $M_\pi $ and $m_N$) in the final formulae. This replacement
(which will take place in all orders $T_{(n)}$) just means that after
writing $F=F_\pi -\delta F_\pi $, etc. and expanding into a power series in
the deltas ($\delta F_\pi $, etc.), we keep in the given order of the chiral
expansion only the lowest order in deltas, higher orders in deltas are
incorporated into higher orders of the chiral expansion. (The reason for
this standard procedure is that it guarantees exactly the same form of $%
T_{(1)}$ in tree calculation, one-loop calculation, etc., and an analogous
statement is true for any higher $T_{(n)}.$)

$T_{(2)}$ is given by the tree graphs with one vertex from
$\widehat{{\cal L}}_{\pi N}^{(2)}$ (labelled as "2-nd" in Figures 1--3)
and all the other vertices from $\widehat{{\cal L}}_{\pi N}^{(1)}$ (Fig. 2). 
Again the bare coupling constants and masses are
replaced by their physical values.

Finally, $T_{(3)}$ is given by the sum $T_{(3)}^{\text{tree}}+T_{(3)}^{\text{%
loop}}+T_{(3)}^\delta .$ Here $T_{(3)}^{\text{tree}}$ is given by the tree
graphs with either one vertex from $\widehat{{\cal L}}_{\pi N}^{(3)}$
(labelled as "3-rd" in Figures 1--3) or two
vertices from $\widehat{{\cal L}}_{\pi N}^{(2)},$ and all the other vertices
from $\widehat{{\cal L}}_{\pi N}^{(1)}$ (Fig. 3). The loop contribution $%
T_{(3)}^{\text{loop}}$ is given by one-loop graphs with vertices from $%
\widehat{{\cal L}}_{\pi N}^{(1)}+$ ${\cal L}_{\pi \pi }^{(2)}$ (Fig. 4). As
always, the bare constants are replaced by their physical values. The third
term $T_{(3)}^\delta $ is just the contribution from this replacement in
lower orders.

Both $T_{(3)}^{\text{tree}}$ and $T_{(3)}^{\text{loop}}$ are infinite and
renormalization scale dependent. It is preferable to have the result in the
explicitly finite and renormalization scale independent form, so we shall
shift all the infinite and $\mu $-dependent terms from $T_{(3)}^{\text{loop}%
} $ to $T_{(3)}^{\text{tree}}.$ These shifted terms are polynomials in
external momenta (otherwise they could not be canceled by the counterterms
from $\widehat{{\cal L}}_{\pi N}^{(3)}$ ) and it is reasonable to shift from
$T_{(3)}^{\text{loop}}$ to $T_{(3)}^{\text{tree}}$ also other possible
polynomial terms (in external momenta, including terms with negative powers
of $w$), so that after adding also $T_{(3)}^\delta $ to $T_{(3)}^{\text{tree}%
},$ one has a clear separation of the amplitude into the part $T_{(3)}^{%
\text{pol}},$ polynomial in external momenta plus poles in $w$, and the rest
$T_{(3)}^{\text{uni}},$ which contains imaginary part, required by unitarity.

\unitlength=0.25mm
\special{em:linewidth 0.4pt}
\linethickness{0.4pt}
\begin{picture}(500.00,150.00)
\put(170.00,20.00){\vector(0,1){20.00}}
\put(169.00,20.00){\vector(0,1){20.00}}
\put(170.00,120.00){\line(0,1){20.00}}
\put(169.00,120.00){\line(0,1){20.00}}
\put(170.00,60.00){\line(2,-1){55.00}}
\put(170.00,100.00){\line(2,1){55.00}}
\put(175.00,100.00){\makebox(0,0)[lt]{1-st}}
\put(175.00,60.00){\makebox(0,0)[lb]{1-st}}
\put(200.00,5.00){\makebox(0,0)[cc]{(a)}}
\put(170.00,40.00){\vector(0,1){40.00}}
\put(170.00,80.00){\vector(0,1){40.00}}
\put(169.00,40.00){\vector(0,1){40.00}}
\put(169.00,80.00){\vector(0,1){40.00}}

\put(320.00,20.00){\vector(0,1){30.00}}
\put(320.00,50.00){\vector(0,1){60.00}}
\put(319.00,20.00){\vector(0,1){30.00}}
\put(319.00,50.00){\vector(0,1){60.00}}
\put(320.00,110.00){\line(0,1){30.00}}
\put(319.00,110.00){\line(0,1){30.00}}
\put(320.00,80.00){\line(1,1){55.00}}
\put(320.00,80.00){\line(1,-1){55.00}}
\put(330.00,80.00){\makebox(0,0)[lc]{1-st}}
\put(350.00,5.00){\makebox(0,0)[cc]{(b)}}
\end{picture}

\begin{quotation}
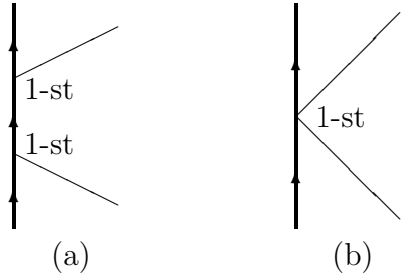
\noindent {\bf Fig.1}{\em \ }{\sl Feynman diagrams for }$\pi ${\sl N
scattering contributing to the first chiral order. Double and single lines
correspond to nucleons and pions, respectively. Crossed diagrams are not
shown.}
\end{quotation}

\smallskip\ \

\unitlength=0.25mm
\special{em:linewidth 0.4pt}
\linethickness{0.4pt}
\begin{picture}(550.00,150.00)
\put(20.00,20.00){\vector(0,1){20.00}}
\put(19.00,20.00){\vector(0,1){20.00}}
\put(20.00,120.00){\line(0,1){20.00}}
\put(19.00,120.00){\line(0,1){20.00}}
\put(20.00,60.00){\line(2,-1){55.00}}
\put(20.00,100.00){\line(2,1){55.00}}
\put(25.00,100.00){\makebox(0,0)[lt]{1-st}}
\put(25.00,60.00){\makebox(0,0)[lb]{2-nd}}
\put(50.00,5.00){\makebox(0,0)[cc]{(a)}}
\put(20.00,40.00){\vector(0,1){40.00}}
\put(20.00,80.00){\vector(0,1){40.00}}
\put(19.00,40.00){\vector(0,1){40.00}}
\put(19.00,80.00){\vector(0,1){40.00}}

\put(170.00,20.00){\vector(0,1){20.00}}
\put(169.00,20.00){\vector(0,1){20.00}}
\put(170.00,120.00){\line(0,1){20.00}}
\put(169.00,120.00){\line(0,1){20.00}}
\put(170.00,60.00){\line(2,-1){55.00}}
\put(170.00,100.00){\line(2,1){55.00}}
\put(175.00,100.00){\makebox(0,0)[lt]{2-nd}}
\put(175.00,60.00){\makebox(0,0)[lb]{1-st}}
\put(200.00,5.00){\makebox(0,0)[cc]{(b)}}
\put(170.00,40.00){\vector(0,1){40.00}}
\put(170.00,80.00){\vector(0,1){40.00}}
\put(169.00,40.00){\vector(0,1){40.00}}
\put(169.00,80.00){\vector(0,1){40.00}}

\put(320.00,20.00){\vector(0,1){20.00}}
\put(319.00,20.00){\vector(0,1){20.00}}
\put(320.00,120.00){\line(0,1){20.00}}
\put(319.00,120.00){\line(0,1){20.00}}
\put(320.00,60.00){\line(2,-1){55.00}}
\put(320.00,100.00){\line(2,1){55.00}}
\put(325.00,100.00){\makebox(0,0)[lt]{1-st}}
\put(325.00,60.00){\makebox(0,0)[lb]{1-st}}
\put(350.00,5.00){\makebox(0,0)[cc]{(c)}}
\put(320.00,80.00){\vector(0,1){40.00}}
\put(319.00,80.00){\vector(0,1){40.00}}
\put(320.00,40.00){\line(0,1){40.00}}
\put(319.00,40.00){\line(0,1){40.00}}
\put(319.50,80.00){\circle*{7.00}}

\put(470.00,20.00){\vector(0,1){30.00}}
\put(470.00,50.00){\vector(0,1){60.00}}
\put(469.00,20.00){\vector(0,1){30.00}}
\put(469.00,50.00){\vector(0,1){60.00}}
\put(470.00,110.00){\line(0,1){30.00}}
\put(469.00,110.00){\line(0,1){30.00}}
\put(470.00,80.00){\line(1,1){55.00}}
\put(470.00,80.00){\line(1,-1){55.00}}
\put(480.00,80.00){\makebox(0,0)[lc]{2-nd}}
\put(500.00,5.00){\makebox(0,0)[cc]{(d)}}
\end{picture}

\begin{quotation}
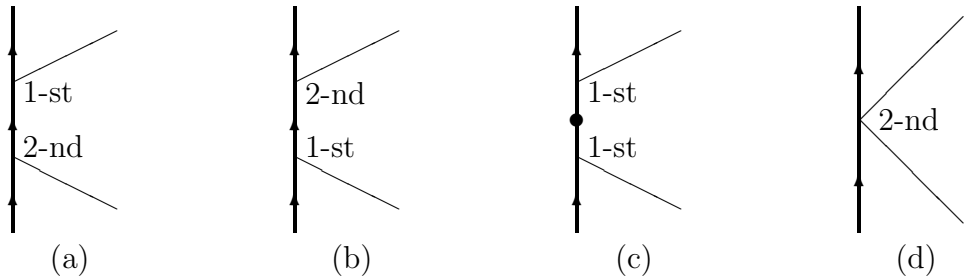
\noindent {\bf Fig.2 }{\sl Diagrams contributing to the second chiral order.
Black circle is the second order counterterm.
Crossed diagrams are not shown.}
\end{quotation}

\smallskip\

\unitlength=0.25mm
\special{em:linewidth 0.4pt}
\linethickness{0.4pt}
\begin{picture}(550.00,150.00)
\put(20.00,20.00){\vector(0,1){20.00}}
\put(19.00,20.00){\vector(0,1){20.00}}
\put(20.00,120.00){\line(0,1){20.00}}
\put(19.00,120.00){\line(0,1){20.00}}
\put(20.00,60.00){\line(2,-1){55.00}}
\put(20.00,100.00){\line(2,1){55.00}}
\put(25.00,100.00){\makebox(0,0)[lt]{1-st}}
\put(25.00,60.00){\makebox(0,0)[lb]{3-rd}}
\put(50.00,5.00){\makebox(0,0)[cc]{(a)}}
\put(20.00,40.00){\vector(0,1){40.00}}
\put(20.00,80.00){\vector(0,1){40.00}}
\put(19.00,40.00){\vector(0,1){40.00}}
\put(19.00,80.00){\vector(0,1){40.00}}

\put(170.00,20.00){\vector(0,1){20.00}}
\put(169.00,20.00){\vector(0,1){20.00}}
\put(170.00,120.00){\line(0,1){20.00}}
\put(169.00,120.00){\line(0,1){20.00}}
\put(170.00,60.00){\line(2,-1){55.00}}
\put(170.00,100.00){\line(2,1){55.00}}
\put(175.00,100.00){\makebox(0,0)[lt]{3-rd}}
\put(175.00,60.00){\makebox(0,0)[lb]{1-st}}
\put(200.00,5.00){\makebox(0,0)[cc]{(b)}}
\put(170.00,40.00){\vector(0,1){40.00}}
\put(170.00,80.00){\vector(0,1){40.00}}
\put(169.00,40.00){\vector(0,1){40.00}}
\put(169.00,80.00){\vector(0,1){40.00}}

\put(320.00,20.00){\vector(0,1){20.00}}
\put(319.00,20.00){\vector(0,1){20.00}}
\put(320.00,120.00){\line(0,1){20.00}}
\put(319.00,120.00){\line(0,1){20.00}}
\put(320.00,60.00){\line(2,-1){55.00}}
\put(320.00,100.00){\line(2,1){55.00}}
\put(325.00,100.00){\makebox(0,0)[lt]{2-nd}}
\put(325.00,60.00){\makebox(0,0)[lb]{2-nd}}
\put(350.00,5.00){\makebox(0,0)[cc]{(c)}}
\put(320.00,40.00){\vector(0,1){40.00}}
\put(320.00,80.00){\vector(0,1){40.00}}
\put(319.00,40.00){\vector(0,1){40.00}}
\put(319.00,80.00){\vector(0,1){40.00}}

\put(470.00,20.00){\vector(0,1){30.00}}
\put(470.00,50.00){\vector(0,1){60.00}}
\put(469.00,20.00){\vector(0,1){30.00}}
\put(469.00,50.00){\vector(0,1){60.00}}
\put(470.00,110.00){\line(0,1){30.00}}
\put(469.00,110.00){\line(0,1){30.00}}
\put(470.00,80.00){\line(1,1){55.00}}
\put(470.00,80.00){\line(1,-1){55.00}}
\put(480.00,80.00){\makebox(0,0)[lc]{3-rd}}
\put(500.00,5.00){\makebox(0,0)[cc]{(d)}}
\end{picture}

\smallskip\

\unitlength=0.25mm
\special{em:linewidth 0.4pt}
\linethickness{0.4pt}
\begin{picture}(500.00,150.00)

\put(90.00,20.00){\vector(0,1){20.00}}
\put(89.00,20.00){\vector(0,1){20.00}}
\put(90.00,120.00){\line(0,1){20.00}}
\put(89.00,120.00){\line(0,1){20.00}}
\put(90.00,60.00){\line(2,-1){55.00}}
\put(90.00,100.00){\line(2,1){55.00}}
\put(95.00,100.00){\makebox(0,0)[lt]{1-st}}
\put(95.00,60.00){\makebox(0,0)[lb]{2-nd}}
\put(120.00,5.00){\makebox(0,0)[cc]{(e)}}
\put(90.00,40.00){\line(0,1){40.00}}
\put(90.00,80.00){\vector(0,1){40.00}}
\put(89.00,40.00){\line(0,1){40.00}}
\put(89.00,80.00){\vector(0,1){40.00}}
\put(89.50,80.00){\circle*{7.00}}

\put(240.00,20.00){\vector(0,1){20.00}}
\put(239.00,20.00){\vector(0,1){20.00}}
\put(240.00,120.00){\line(0,1){20.00}}
\put(239.00,120.00){\line(0,1){20.00}}
\put(240.00,60.00){\line(2,-1){55.00}}
\put(240.00,100.00){\line(2,1){55.00}}
\put(245.00,100.00){\makebox(0,0)[lt]{2-nd}}
\put(245.00,60.00){\makebox(0,0)[lb]{1-st}}
\put(270.00,5.00){\makebox(0,0)[cc]{(f)}}
\put(240.00,40.00){\line(0,1){40.00}}
\put(240.00,80.00){\vector(0,1){40.00}}
\put(239.00,40.00){\line(0,1){40.00}}
\put(239.00,80.00){\vector(0,1){40.00}}
\put(239.50,80.00){\circle*{7.00}}

\put(375.00,20.00){\vector(0,1){20.00}}
\put(374.00,20.00){\vector(0,1){20.00}}
\put(375.00,120.00){\line(0,1){20.00}}
\put(374.00,120.00){\line(0,1){20.00}}
\put(375.00,60.00){\line(2,-1){55.00}}
\put(375.00,100.00){\line(2,1){55.00}}
\put(380.00,100.00){\makebox(0,0)[lt]{1-st}}
\put(380.00,60.00){\makebox(0,0)[lb]{1-st}}
\put(405.00,5.00){\makebox(0,0)[cc]{(g)}}
\put(375.00,40.00){\line(0,1){40.00}}
\put(375.00,80.00){\vector(0,1){40.00}}
\put(374.00,40.00){\line(0,1){40.00}}
\put(374.00,80.00){\vector(0,1){40.00}}
\put(374.50,70.00){\circle*{7.00}}
\put(374.50,90.00){\circle*{7.00}}

\end{picture}

\begin{quotation}
\noindent {\bf Fig.3 }{\sl Tree diagrams contributing to the third chiral
order. Crossed diagrams are not shown.}
\end{quotation}

\newpage\

\unitlength=0.25mm
\special{em:linewidth 0.4pt}
\linethickness{0.4pt}
\begin{picture}(550.00,150.00)

\put(20.00,20.00){\line(0,1){120.00}}
\put(19.00,20.00){\line(0,1){120.00}}
\put(20.00,60.00){\line(2,-1){55.00}}
\put(20.00,100.00){\line(2,1){55.00}}
\put(50.00,5.00){\makebox(0,0)[cc]{(a)}}
\put(19.00,90.00){\oval(20.00,20.00)[l]}

\put(170.00,20.00){\line(0,1){120.00}}
\put(169.00,20.00){\line(0,1){120.00}}
\put(170.00,60.00){\line(2,-1){55.00}}
\put(170.00,100.00){\line(2,1){55.00}}
\put(200.00,5.00){\makebox(0,0)[cc]{(b)}}
\put(169.00,70.00){\oval(20.00,20.00)[l]}

\put(320.00,20.00){\line(0,1){120.00}}
\put(319.00,20.00){\line(0,1){120.00}}
\put(320.00,60.00){\line(2,-1){55.00}}
\put(320.00,100.00){\line(2,1){55.00}}
\put(350.00,5.00){\makebox(0,0)[cc]{(c)}}
\put(319.00,100.00){\oval(40.00,40.00)[l]}

\put(470.00,20.00){\line(0,1){120.00}}
\put(469.00,20.00){\line(0,1){120.00}}
\put(470.00,60.00){\line(2,-1){55.00}}
\put(470.00,100.00){\line(2,1){55.00}}
\put(500.00,5.00){\makebox(0,0)[cc]{(d)}}
\put(469.00,60.00){\oval(40.00,40.00)[l]}

\end{picture}

\unitlength=0.25mm
\special{em:linewidth 0.4pt}
\linethickness{0.4pt}
\begin{picture}(550.00,150.00)

\put(20.00,20.00){\line(0,1){120.00}}
\put(19.00,20.00){\line(0,1){120.00}}
\put(20.00,60.00){\line(2,-1){55.00}}
\put(20.00,100.00){\line(2,1){55.00}}
\put(50.00,5.00){\makebox(0,0)[cc]{(e)}}
\put(19.00,80.00){\oval(20.00,20.00)[l]}

\put(170.00,20.00){\line(0,1){120.00}}
\put(169.00,20.00){\line(0,1){120.00}}
\put(170.00,60.00){\line(2,-1){55.00}}
\put(170.00,100.00){\line(2,1){55.00}}
\put(200.00,5.00){\makebox(0,0)[cc]{(f)}}
\put(169.00,80.00){\oval(40.00,40.00)[l]}

\put(320.00,20.00){\line(0,1){120.00}}
\put(319.00,20.00){\line(0,1){120.00}}
\put(320.00,60.00){\line(2,-1){55.00}}
\put(320.00,100.00){\line(2,1){55.00}}
\put(350.00,5.00){\makebox(0,0)[cc]{(g)}}
\put(319.00,90.00){\oval(40.00,60.00)[l]}

\put(470.00,20.00){\line(0,1){120.00}}
\put(469.00,20.00){\line(0,1){120.00}}
\put(470.00,60.00){\line(2,-1){55.00}}
\put(470.00,100.00){\line(2,1){55.00}}
\put(500.00,5.00){\makebox(0,0)[cc]{(h)}}
\put(469.00,70.00){\oval(40.00,60.00)[l]}

\end{picture}

\unitlength=0.25mm
\special{em:linewidth 0.4pt}
\linethickness{0.4pt}
\begin{picture}(550.00,150.00)

\put(20.00,20.00){\line(0,1){120.00}}
\put(19.00,20.00){\line(0,1){120.00}}
\put(20.00,60.00){\line(2,-1){55.00}}
\put(20.00,100.00){\line(2,1){55.00}}
\put(50.00,5.00){\makebox(0,0)[cc]{(i)}}
\put(19.00,80.00){\oval(40.00,80.00)[l]}

\put(170.00,20.00){\line(0,1){120.00}}
\put(169.00,20.00){\line(0,1){120.00}}
\put(210.00,80.00){\line(1,1){55.00}}
\put(210.00,80.00){\line(1,-1){55.00}}
\put(200.00,5.00){\makebox(0,0)[cc]{(k)}}
\put(190.00,80.00){\circle{40.00}}

\put(320.00,20.00){\line(0,1){120.00}}
\put(319.00,20.00){\line(0,1){120.00}}
\put(350.00,80.00){\line(1,1){55.00}}
\put(350.00,80.00){\line(1,-1){55.00}}
\put(350.00,5.00){\makebox(0,0)[cc]{(l)}}
\put(320.00,80.00){\oval(60.00,40.00)[r]}

\put(470.00,20.00){\line(0,1){120.00}}
\put(469.00,20.00){\line(0,1){120.00}}
\put(470.00,80.00){\line(1,1){55.00}}
\put(470.00,80.00){\line(1,-1){55.00}}
\put(500.00,5.00){\makebox(0,0)[cc]{(m)}}
\put(469.00,80.00){\oval(40.00,40.00)[l]}

\end{picture}

\unitlength=0.25mm
\special{em:linewidth 0.4pt}
\linethickness{0.4pt}
\begin{picture}(550.00,150.00)

\put(20.00,20.00){\line(0,1){120.00}}
\put(19.00,20.00){\line(0,1){120.00}}
\put(20.00,60.00){\line(2,-1){55.00}}
\put(20.00,100.00){\line(2,1){55.00}}
\put(50.00,5.00){\makebox(0,0)[cc]{(n)}}
\put(19.00,115.00){\oval(30.00,30.00)[l]}

\put(170.00,20.00){\line(0,1){120.00}}
\put(169.00,20.00){\line(0,1){120.00}}
\put(170.00,60.00){\line(2,-1){55.00}}
\put(170.00,100.00){\line(2,1){55.00}}
\put(200.00,5.00){\makebox(0,0)[cc]{(o)}}
\put(169.00,45.00){\oval(30.00,30.00)[l]}

\put(320.00,20.00){\line(0,1){120.00}}
\put(319.00,20.00){\line(0,1){120.00}}
\put(320.00,60.00){\line(2,-1){55.00}}
\put(320.00,100.00){\line(2,1){55.00}}
\put(350.00,5.00){\makebox(0,0)[cc]{(p)}}
\put(304.00,100.00){\circle{30.00}}

\put(470.00,20.00){\line(0,1){120.00}}
\put(469.00,20.00){\line(0,1){120.00}}
\put(470.00,60.00){\line(2,-1){55.00}}
\put(470.00,100.00){\line(2,1){55.00}}
\put(500.00,5.00){\makebox(0,0)[cc]{(r)}}
\put(454.00,60.00){\circle{30.00}}

\end{picture}

\unitlength=0.25mm
\special{em:linewidth 0.4pt}
\linethickness{0.4pt}
\begin{picture}(550.00,150.00)

\put(20.00,20.00){\line(0,1){120.00}}
\put(19.00,20.00){\line(0,1){120.00}}
\put(20.00,60.00){\line(2,-1){55.00}}
\put(20.00,100.00){\line(2,1){55.00}}
\put(50.00,5.00){\makebox(0,0)[cc]{(s)}}
\put(4.00,80.00){\circle{30.00}}

\put(170.00,20.00){\line(0,1){120.00}}
\put(169.00,20.00){\line(0,1){120.00}}
\put(170.00,80.00){\line(1,1){55.00}}
\put(170.00,80.00){\line(1,-1){55.00}}
\put(200.00,5.00){\makebox(0,0)[cc]{(t)}}
\put(169.00,95.00){\oval(30.00,30.00)[l]}

\put(320.00,20.00){\line(0,1){120.00}}
\put(319.00,20.00){\line(0,1){120.00}}
\put(320.00,80.00){\line(1,1){55.00}}
\put(320.00,80.00){\line(1,-1){55.00}}
\put(350.00,5.00){\makebox(0,0)[cc]{(u)}}
\put(319.00,65.00){\oval(30.00,30.00)[l]}

\put(470.00,20.00){\line(0,1){120.00}}
\put(469.00,20.00){\line(0,1){120.00}}
\put(470.00,80.00){\line(1,1){55.00}}
\put(470.00,80.00){\line(1,-1){55.00}}
\put(500.00,5.00){\makebox(0,0)[cc]{(v)}}
\put(454.00,80.00){\circle{30.00}}

\end{picture}

\bigskip\

\begin{quotation}
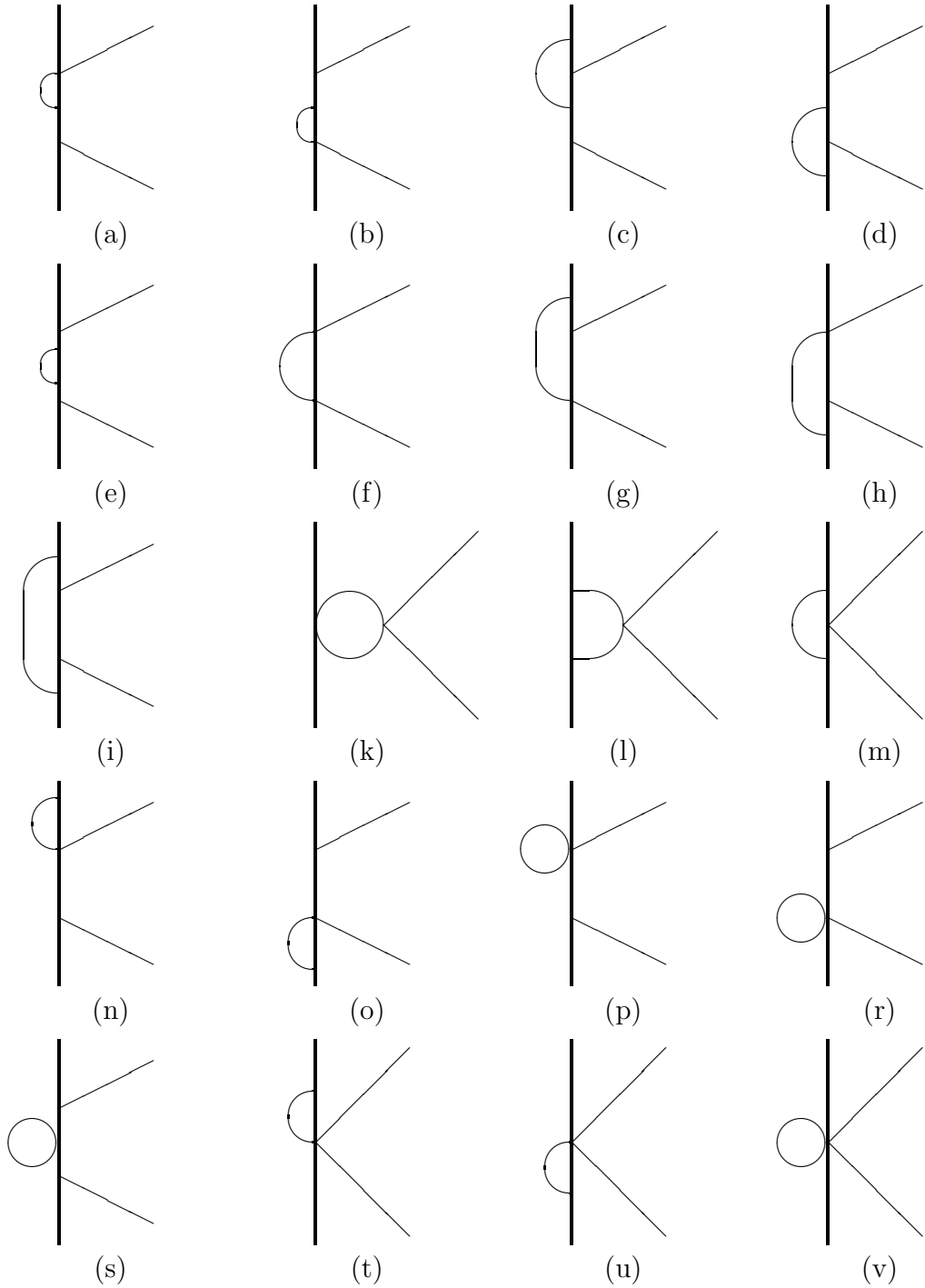
\noindent {\bf Fig.4 }{\sl 1-loop diagrams contributing to the third chiral
order. Crossed diagrams are not shown.}
\end{quotation}

We shall therefore present results at the third order in the form
\begin{equation}
\label{T3}T_{(3)}=T_{(3)}^{\text{pol}}+T_{(3)}^{\text{uni}}\quad .
\end{equation}
$T_{(3)}^{\text{uni}}$ is made finite and scale independent by hand, $%
T_{(3)}^{\text{pol}}$ by the cancellations of divergent and scale dependent
terms from loops and counterterms. The counterterms in the third-order
Lagrangian (Table \ref{Tab O}) were completely fixed by the general
renormalization of the generating functional \cite{E94}, and so this
cancellation provides us with a nontrivial check of the calculations.

Calculation of $T$ from (\ref{LMB}) is in principle straightforward. Here we
present just the final result, details are postponed to appendix \ref
{diagrams}. For loop functions the following notation is used 
\begin{equation}
\label{IJ uni}
\begin{array}{lcl}
J_{\pm }^{\text{uni}}(w) & = & J_0^{
\text{uni}}(w)\pm J_0^{\text{uni}}(-w) \\ J_0^{\text{uni}}(w) & = & 
J_0(w)+4wL(\mu )-\frac w{8\pi ^2}\left( 1-2\ln \frac M\mu \right) \\ 
I_0^{\text{uni}}(t) & = & I_0(t)+2L(\mu )-\frac 1{16\pi ^2}\left( 1-2\ln
\frac M\mu \right) 
\end{array}
\end{equation}
and the loop integrals $I_0,$ $J_0,$ and $K_0$ (which will appear later) are
displayed in appendix \ref{loops}. All the loop integrals in the amplitude
are understood with the bare pion mass $M$ replaced by the physical mass $%
M_\pi $.

Let us remark that two of the LECs from Table \ref{Tab O} --- namely $b_4$
and $b_{17}$ --- do not appear in the result. The reason for $b_4$ is that
the term containing this LEC is of the form $w\,p\cdot (q+q^{\prime })$
which is of the fourth chiral order, since $p$ itself is of the second order
for the on-shell nucleon. The reason for $b_{17}$ is, on the other hand,
cancellation (in the third order) of different terms containing this LEC.
All the other LECs from Table \ref{Tab O} are present in the result given
below\footnote{%
For sake of completeness let us note that when taking into account also
isospin breaking effects, three more LECs would appear, namely $a_4,$ $%
b_{18} $ and $b_{20}.$}:

\begin{equation}
\label{alf p}
\begin{array}{rcl}
\alpha _{(1)}^{+} & = & \frac{g_A^2}{4F_\pi ^2}\left( \frac 1w-\frac
1{w^{\prime }}\right) \left( q\cdot q^{\prime }-ww^{\prime }\right) \\  &  &
\\ 
\alpha _{(2)}^{+} & = & \frac 4{m_NF_\pi ^2}\left( a_1q\cdot q^{\prime
}+a_2\;ww^{\prime }-a_3M_\pi ^2\right) \\
& + & \frac{g_A^2}{8m_NF_\pi ^2}\left[ \left( \frac 1{w^2}+\frac
1{w^{^{\prime }2}}\right) \;\left( ww^{\prime }q\cdot q^{\prime }-M_\pi
^2ww^{\prime }-M_\pi ^2q\cdot q^{\prime }-w^2w^{^{\prime }2}\right) \right.
\\  & + & \left. 6ww^{\prime }-2q\cdot q^{\prime }\right] \\
&  &  \\
\alpha _{(3)}^{+\text{ pol}} & = & \frac{g_A^2M_\pi }{32\pi F_\pi ^4}\left(
4q\cdot q^{\prime }-3M_\pi ^2\right) +\frac{g_A^4M_\pi ^3}{12\pi F_\pi ^4}%
\frac 1{w^2}\left( q\cdot q^{\prime }-w^2\right) \\  &  &  \\
\alpha _{(3)}^{+\text{ uni}} & = & \frac 1{2F_\pi ^4}w^2J_{+}^{
\text{uni}}(w)-\frac{g_A^2}{8F_\pi ^4}\left( 2M_\pi ^2-t\right) \left( M_\pi
^2-2t\right) K_0(0,t) \\  & + & \frac{g_A^4}{3F_\pi ^4}\frac 1{w^2}\left(
q\cdot q^{\prime }-w^2\right) \left( M_\pi ^2-w^2\right) J_{+}^{\text{uni}%
}(w)
\end{array}
\end{equation}

\medskip\

\begin{equation}
\label{bet p}
\begin{array}{rcl}
\beta _{(1)}^{+} & = & \frac{g_A^2}{2F_\pi ^2}\left( \frac 1w+\frac
1{w^{\prime }}\right) \\  &  &  \\
\beta _{(2)}^{+} & = & \frac{g_A^2}{4m_NF_\pi ^2}\left( \frac 1{w^2}-\frac
1{w^{\prime 2}}\right) \left( ww^{\prime }-M_\pi ^2\right) \\  &  &  \\
\beta _{(3)}^{+\text{ pol}} & = & \frac 1{4\pi ^2F_\pi ^4}\left[ \left(
b_{16}-
\widetilde{b}_{15}\right) w-b_{19}\frac{g_AM_\pi ^2}w\right] \\  & + & \frac{%
g_A^2M_\pi ^4}{4m_N^2F_\pi ^2}\frac 1{w^3}+\frac{g_A^4}{72\pi ^2F_\pi ^4}%
\frac 1w(w^2+6M_\pi ^2)\qquad \quad \\  &  &  \\
\beta _{(3)}^{+\text{ uni}} & = & \frac{g_A^4}{6F_\pi ^4}\frac 1{w^2}(M_\pi
^2-w^2)J_{-}^{\text{uni}}(w)
\end{array}
\end{equation}

\medskip\

\begin{equation}
\label{alf m}
\begin{array}{rcl}
\alpha _{(1)}^{-} & = & \frac{g_A^2}{4F_\pi ^2}\left( \frac 1w+\frac
1{w^{\prime }}\right) \left( q\cdot q^{\prime }-ww^{\prime }\right) +\frac
1{4F_\pi^2}(w+w^{\prime }) \\  &  &  \\
\alpha _{(2)}^{-} & = & \frac{g_A^2}{8m_NF_\pi ^2}\;\left( \frac
1{w^2}-\frac 1{w^{\prime 2}}\right) \left( ww^{\prime }q\cdot q^{\prime
}-M_\pi ^2ww^{\prime }-M_\pi ^2q\cdot q^{\prime }-w^2w^{^{\prime }2}\right)
\\  &  &  \\
\alpha _{(3)}^{-\text{ pol}} & = & \frac 1{4\pi ^2F_\pi ^4}w\left[ (
\widetilde{b}_1+\widetilde{b}_2)\;q\cdot q^{\prime }+\widetilde{b}_3w^2+2%
\widetilde{b}_6M_\pi ^2-b_{19}\frac{g_AM_\pi ^2}{2w^2}\left( q\cdot
q^{\prime }-w^2\right) \right] \\  & + & \frac{g_A^2M_\pi ^2}{8m_N^2F_\pi ^2}%
\frac 1w\left\{ \left( q\cdot q^{\prime }-w^2\right) \left[ 1+\frac{M_\pi ^2%
}{w^2}+\frac{w^2}{M_\pi ^2}\right] +3\left( M_\pi ^2-q\cdot q^{\prime
}\right) \right\} \\  & + & \frac 1{288\pi ^2F_\pi ^4}w\left[ 18w^2-\left(
7+11g_A^2\right) M_\pi ^2-\left( 5+13g_A^2\right) q\cdot q^{\prime }\right]
\\
& + & \frac{g_A^4}{144\pi ^2F_\pi ^4}\frac 1w\left( q\cdot q^{\prime
}-w^2\right) \left( 6M_\pi ^2-5w^2\right) +\frac 2{m_N^2F_\pi ^2}w\left(
M_\pi ^2a_3+\frac 1{32}t\right) \\  &  &  \\
\alpha _{(3)}^{-\text{ uni}} & = & \frac 1{12F_\pi ^4}\left[ 3w^2+g_A^4\frac
1{w^2}(q\cdot q^{\prime }-w^2)(M_\pi ^2-w^2)\right] J_{-}^{
\text{uni}}(w) \\  & - & \frac 1{3F_\pi ^4}w\left[ \left( M_\pi ^2-\frac
14t\right) +2g_A^2\left( M_\pi ^2-\frac 58t\right) \right] I_0^{\text{uni}%
}(t)
\end{array}
\end{equation}

\medskip\
\begin{equation}
\label{bet m}
\begin{array}{rcl}
\beta _{(1)}^{-} & = & \frac{g_A^2}{2F_\pi ^2}\left( \frac 1w-\frac
1{w^{\prime }}\right) \\  &  &  \\
\beta _{(2)}^{-} & = & -\frac 2{m_NF_\pi ^2}a_5+
\frac{g_A^2}{4m_NF_\pi ^2}\left[ \left( \frac 1{w^2}+\frac 1{w^{\prime
2}}\right) \left( ww^{\prime }-M_\pi ^2\right) -2\right] \qquad \qquad \quad
\\  &  &  \\
\beta _{(3)}^{-\text{ pol}} & = & \frac{g_A^2M_\pi }{16\pi F_\pi ^4}+\frac{%
g_A^4M_\pi ^3}{12\pi F_\pi ^4}\frac 1{w^2} \\  &  &  \\ 
\beta _{(3)}^{-\text{ uni}} & = & \frac{g_A^2}{4F_\pi ^4}\left( t-4M_\pi
^2\right) K_0(0,t)+\frac{g_A^4}{3F_\pi ^4}\frac 1{w^2}(M_\pi ^2-w^2)J_{+}^{%
\text{uni}}(w)\quad . 
\end{array}
\end{equation}

\newpage

\section{Comparison with data}

\subsection{Partial waves and threshold parameters}

To compare with experiment, it is convenient to express the results in terms
of $A^{\pm }$ and $B^{\pm },$ defined by the standard parametrization of the
elastic $\pi $N scattering amplitude
\begin{equation}
\label{TpN}T_{\pi N}^{\pm }=\overline{u}(mv+p^{^{\prime }},\sigma ^{^{\prime
}})\left[ A^{\pm }(s,t,u)+B^{\pm }(s,t,u)\frac{\NEG q+\NEG q^{^{\prime }}}%
2\right] u(mv+p,\sigma )\quad .
\end{equation}
Starting from (\ref{uN}) one derives (after some algebra ) following
relations between $\alpha ^{\pm },$ $\beta ^{\pm }$ and $A^{\pm },$ $B^{\pm }
$ (see also Ref. \cite{EM97})
\begin{equation}
\label{AB}
\begin{array}{rcl}
A^{\pm } & = & \left( \alpha ^{\pm }+
\frac{s-u}4\beta ^{\pm }\right)  \\ B^{\pm } & = & \left( -m_N+\frac
t{4m_N}\right) \beta _{}^{\pm }\quad .
\end{array}
\end{equation}

Experimental information about elastic $\pi $N-scattering is usually given
in terms of partial waves. For low energies, partial waves are characterized
by a small number of threshold parameters such as scattering lengths,
volumes, effective ranges, etc. These threshold parameters are not directly
measurable, they can be$\;$extrapolated, however, from the experimental data
above threshold \cite{H83}\cite{H96}. Since the closer to threshold one is,
the better CHPT is expected to work, these parameters seem to be the most
suitable quantities to calculate and to compare with the (extrapolated)
experimental values.

Partial wave amplitudes for the scattering of a spinless particle on a spin $%
\frac 12$ particle are given by \cite{H83} 
\begin{equation}
\label{fpm}
\begin{array}{ccc}
f_{l\pm }^{\pm }(s) & = & \frac 1{16\pi 
\sqrt{s}}\int_{-1}^1\{(E+m_N)[A^{\pm }(s,\theta )+(\sqrt{s}-m_N)B^{\pm
}(s,\theta )]P_l(\cos \theta ) \\  & - & (E-m_N)[A^{\pm }(s,\theta )-(\sqrt{s%
}+m_N)B^{\pm }(s,\theta )]P_{l\pm 1}(\cos \theta )\}\;d\cos \theta
\end{array}
\end{equation}
where $E$ is the nucleon energy in CMS, and $\theta $ is the scattering angle in
CMS. The low-energy behavior of the partial wave amplitudes is characterized
by the threshold parameters $a_{l\pm }^{\pm }$, $b_{l\pm }^{\pm }$ defined
by the expansion
\begin{equation}
\label{wave exp}\text{Re }f_{l\pm }^{\pm }(s)=a_{l\pm }^{\pm }k^{2l}+b_{l\pm
}^{\pm }k^{2l+2}+\ldots
\end{equation}
where $\stackrel{\rightarrow }{k}$ is the 3-momentum of a particle in CMS.

Calculation of these threshold parameters from the scattering amplitude (\ref
{alf p}--\ref{bet m}) is in principle straightforward, results for the first
four partial waves are given in appendix \ref{lengths}.

\subsection{Threshold parameters independent of LECs $a_i,\ b_j$}

The scattering amplitude (\ref{alf p}--\ref{bet m}), and therefore also
the sixteen threshold parameters given in appendix \ref{lengths},
contain four LECs of the second-order $\pi $N Lagrangian ($a_1,$ $a_2,$ $%
a_3, $ $a_5)$ and five linear combinations of LECs of the third-order $\pi $%
N Lagrangian ($\widetilde{b}_1+\widetilde{b}_2,$ $\widetilde{b}_3,$ $%
\widetilde{b}_6,$ $b_{16}-\widetilde{b}_{15},$ $b_{19}$). However, six out
of these sixteen threshold parameters do not depend on any of these LECs.
So we have at our disposal six parameter-free (where by parameters the LECs
$a_i$ and $b_j$ are understood) HBCHPT predictions.

\begin{table} \centering
  \begin{tabular}{|c|c|c|c|c|} \hline
     & {up to \bf $\QTR{cal}{O}(p)$}  &  {up to \bf $\QTR{cal}{O}(p^2)$}  &
        {up to \bf $\QTR{cal}{O}(p^3)$} & {\bf exp.} \\ \hline
    $a^+_{2+} $  &  $\ -55 $ &  $\ -59 $  &  $\ -36 $      &  $\  -36 \pm 7\ $ \\ \hline
    $a^-_{2+} $  &  $\quad \ 55 $ &  $\quad \ 59$  &  $\quad  \ 56 $      &  $\quad 64 \pm 3\ $ \\ \hline
    $a^+_{3+} $  &  $\quad 180 $ &  $\quad 210 $  &  $\quad 280 $      &  $\quad 440 \pm 140 $ \\ \hline
    $a^+_{3-} $  &  $\ -30 $  &  $\ -31 $  &  $\quad \ 31$      &  $\quad 160 \pm 120$ \\ \hline
    $a^-_{3+} $  &  $-180 $      &  $-210 $  &  $-210 $      &  $-260 \pm 20\ $ \\ \hline
    $a^-_{3-} $  &  $\quad \ 30 $      &  $\quad \  34$  &  $\quad  \ 57 $      &  $\quad100 \pm 20\ $ \\ \hline
  \end{tabular}
  \caption{Comparison of the HBCHPT results for two D-wave and four F-wave
threshold parameters up to the first, second and third order, with (extrapolated)
experimental values. Units for D-wave are $GeV^{-5},$ for F-wave they
are $GeV^{-7}.$
\label{Tab thres1}}
\end{table}

To compare them with experimental data, one needs some consistent set of
extrapolated threshold parameters, for which we take Ref. \cite{K80}. This
comparison, which is shown in Table \ref{Tab thres1}, is neither too
impressive, nor discouraging.\footnote{%
One should be aware that experimental errors in \cite{K80} are
underestimated, since they refer ''only to that part which can be estimated
from deviations from the internal consistency of the method''.} Two main
lessons can be drawn from Table \ref{Tab thres1}:

\begin{itemize}
\item  There is a clear improvement, when going from lower to higher orders
of HBCHPT.

\item  Contributions of different orders are comparable, i.e. the chiral
expansion converges slowly.
\end{itemize}

Our results for D- and F-wave threshold parameters are in agreement with
independent calculation of these parameters in \cite{BKM96}. In this work
the Born term was treated in different manner, namely the ''relativistic 1-st
order'' approach (discussed above) was used and results are presented to the
leading (F-waves) and next-to-leading (D-waves) order in powers of
$\mu =M_\pi /m_N$.
Our calculation, on the other hand, is fully based on the HBCHPT Lagrangian.
For every threshold parameter, the two approaches give the same result to the
order in powers of $\mu$  given in \cite{BKM96}, but
there are differences in higher orders of $\mu ,$ leading in some cases to
nonnegligible (but not at all dramatic) differences in numerical values of
the threshold parameters.

\subsection{Remaining threshold parameters and LECs $a_i,\ b_j$}

Let us now turn our attention to the remaining ten threshold parameters. As
mentioned above, they contain nine LECs $a_i,\ b_j$ and/or their linear
combinations. Two of them are present also in HBCHPT results for two other
quantities closely related to $\pi $N scattering. The first one is the
pion-nucleon $\sigma$-term
\begin{equation}
\label{sigma def}\sigma =\frac{1}{2m_N}\left\langle P\left| \widehat{m}(\overline{u}u+%
\overline{d}d)\right| P\right\rangle
\end{equation}
where $P$ is the nucleon 4-momentum, $\widehat{m}=(m_u+m_d)/2$ and $u,$ $d$
are quark fields. The second one is the Goldberger-Treiman discrepancy $%
1 - g_Am_N/F_\pi g_{\pi N} .$ Up to the third order of HBCHPT \cite{BKM95}

\begin{equation}
\label{sigma}\sigma =-\frac{4M_\pi ^2}{m_N}a_3-\frac{9g_A^2M_\pi ^3}{64\pi
F_\pi ^2}
\end{equation}
\begin{equation}
\label{G-T}g_{\pi N}=g_A\frac{m_N}{F_\pi }\left( 1-\frac{M_\pi ^2}{8\pi
^2F_\pi ^2g_A}b_{19}\right) \quad .
\end{equation}

\begin{table} \centering
  \begin{tabular}{|c|c|} \hline
     $a_1$  &  $-2.60 \pm 0.03 $ \\ \hline
     $a_2$  &  $\quad 1.40\ \pm 0.05$  \\ \hline
     $a_3$  &  $-1.00\ \pm 0.06$  \\ \hline
     $a_5$  &  $\quad 3.30 \pm 0.05 $ \\ \hline
     $\widetilde{b}_1+\widetilde{b}_2$      &  $\quad 2.4\ \pm 0.3 $ \\ \hline
     $\widetilde{b}_3$       &  $-2.8\ \pm 0.6 $  \\ \hline
     $\widetilde{b}_6$       & $\quad 1.4\ \pm 0.3 $  \\ \hline
     $b_{16}-\widetilde{b}_{15}$      & $\quad 6.1\ \pm 0.6 $  \\ \hline
     $b_{19}$       & $ -2.4  \ \pm 0.4  $ \\ \hline
  \end{tabular}
  \caption{
The values of LECs from the $\chi ^2$-fit to the 10 $\pi$N
scattering threshold parameters,  the nucleon $\sigma$-term, and the Goldberger-Treiman
discrepancy. Uncertainities refer to the parabolic errors of the MINUIT routine.
\label{Tab LEC}}
\end{table}

Values of the LECs can now be fixed by fitting the HBCHPT results to the
(extrapolated) experimental values of the remaining ten threshold
parameters, $\sigma $-term and $g_{\pi N}$. Results of such a procedure,
using values $M_\pi =138$MeV, $m_N=939$MeV, $F_\pi =93$MeV, $g_A=1.26, \sigma
=45\pm 8$MeV, $g_{\pi N}=13.4\pm 0.1$ and threshold parameters from \cite{K80}, are
presented in Table \ref{Tab LEC}. (Let us recall that values of
$b_{16}-\widetilde{b}_{15}$ and $b_{19}$ depend on the choice of mesonic
Lagrangian ${\cal L}_{\pi \pi }^{(4)}$ \cite{EM96}.)

One should stress that errors given in Table \ref{Tab LEC} are probably too
optimistic, as a consequence of probably too optimistic errors of the used
(extrapolated) experimental values of the threshold parameters. Another
consequence of these probably underestimated errors is the relatively large
value of the $\sigma $-term, $\sigma =59\pm 5\,$MeV (corresponding to $%
a_3=-1.00\pm 0.06$). Enhancing error bars of the threshold parameters leads
to a value of $\sigma $-term closer to the experimental one. Another
possibility, how to avoid a large $\sigma $-term, is to fix $a_3$ directly
from the $\sigma $-term and not to treat it as free parameter in the fitting
procedure. When doing so, it turns out that the third order LECs $b_i$ are
not sensitive to the precise value of the $\sigma $-term. Variation of $a_3$
from $-0.8$ to $-1.0$ (corresponding to variation of $\sigma $ from $45$ to $%
60$ MeV) leaves the third order LECs almost untouched.

On the other hand, the third order LECs are sensitive to the value of $%
g_{\pi N}.$ For $g_{\pi N}=13.0\pm 0.1$ one obtains $\widetilde{b}_1+%
\widetilde{b}_2=3.3\pm 0.3,$ $\widetilde{b}_3=-3.7\pm 0.6,$ $\widetilde{b}%
_6=1.4\pm 0.3,$ $b_{16}-\widetilde{b}_{15}=7.9\pm 0.6$ and $b_{19}=-1.0\pm
0.4$ $.$

The values of the second order LECs $a_i$ are in good agreement with their
recent determination \cite{BKM96} ($a_1=-2.59\pm 0.2,$ $a_2=1.57\pm 0.1,$ $%
a_3=-0.92\pm 0.1,$ $a_5=3.48\pm 0.05$ in our conventions). These values were
obtained from a fit to the $\sigma $-term and eight other observables, four
of them being some linear combinations of the threshold parameters used in
our fit and another four being subthreshold parameters, not used by us. So
the agreement is not a trivial one. As a matter of fact, the quoted
results of \cite{BKM96} were obtained in a fit with enhanced error bars for
threshold and subthreshold parameters (for reasons mentioned above).
Original error bars lead to $a_1=-2.68\pm 0.03,$ $a_2=1.47\pm 0.02,$ $%
a_3=-1.02\pm 0.06,$ in even better agreement with our results.

Once we have pinned down the LECs, we can compare explicitly our results for
the threshold parameters with the values given in \cite{K80}. This is done,
order by order, in Table \ref{Tab thres2}.

\begin{table} \centering
  \begin{tabular}{|c|c|c|c|c|c|} \hline
     & {up to \bf $\QTR{cal}{O}(p)$}  &  {up to \bf $\QTR{cal}{O}(p^2)$}  &
        {up to \bf $\QTR{cal}{O}(p^3)$} & {\bf exp.}  & units\\ \hline
    $a^+_0 $  &  $\quad \ 0 \ $      &  $\quad -0.13 $  &  $\quad-0.07 \pm 0.09$      &  $\quad \ -0.07 \pm 0.01 $ & $GeV^{-1}$ \\ \hline
    $b^+_0 $  &  $\quad \ 0 \ $      &  $\ -24 $  &  $ -13.9 \pm 3.0$      &  $-16.9 \pm 2.5 $ & $GeV^{-3}$  \\ \hline
    $a^-_0 $  &  $\qquad 0.55 $      &  $\quad\ 0.55 $  &  $\quad \ 0.67 \pm 0.10 $      &  $\qquad 0.66 \pm 0.01$ & $GeV^{-1}$  \\ \hline
    $b^-_0 $  &  $\quad \ 7.2$      &  $\quad\ 8.2 $  &  $\quad\ 5.5 \ \pm 6.7 $      &  $\quad 5.1 \pm 2.3 $ & $GeV^{-3}$   \\ \hline
    $a^+_{1+} $  &  $\quad 17.6 $      &  $\quad 51.2 $  &  $\quad 50.4 \pm 1.1 $      &  $\quad 50.5 \pm 0.5 $ & $GeV^{-3}$   \\ \hline
    $a^+_{1-} $  &  $-35.3 $      &  $\  -1.7 $  &  $-21.6 \pm 1.8 $      &  $-21.6 \pm 0.5 $ & $GeV^{-3}$   \\ \hline
    $a^-_{1+} $  &  $-16.9 $      &  $ -29.3 $  &  $-31.0 \pm 0.8 $      &  $-31.0 \pm 0.6 $ & $GeV^{-3}$   \\ \hline
    $a^-_{1-} $  &  $-17.1 $      &  $\quad\ 6.7 $  &  $\ -4.5 \pm 1.0 $      &  $\ -4.4 \pm 0.4 $ & $GeV^{-3}$ \\ \hline
    $a^+_{2-} $  &  $\quad  18.6 $      &  $\quad\ 4.6 $  &  $\quad 31.2 \pm 0.3 $      &  $\quad 44\ \ \pm 7 \ $ & $GeV^{-5}$   \\ \hline
    $a^-_{2-} $  &  $\ -8.8 $      &  $-17.6 $  &  $\ -5.0 \pm 0.2  $      &  $\quad 2\ \ \pm 3 $ & $GeV^{-5}$   \\ \hline
   \end{tabular}
  \caption{HBCHPT predictions for the 10 elastic $\pi$N scattering
threshold parameters up to the first, second and third order, compared with
(extrapolated) experimental values.
Theoretical uncertainties are shown only in the third order.\label{Tab thres2}}
\end{table}

At first sight, there is remarkable agreement between theory and experiment.
This statement is however almost empty, since we have chosen the values of
LECs just to obtain such an agreement. The nontrivial content of this
agreement is that it was achieved with reasonable, i.e. not unnaturally
large, values of LECs.

A criterion for the natural size of LECs is that corresponding counterterm
contributions are not too large when compared to other contributions (loops
and counterterms with fixed coefficients and/or lower order LECs) of the
same order. Of course, such a comparison requires choosing some typical
renormalization scale $\mu .$ Numerical results for $\mu =1$GeV are
presented in Table \ref{Tab comp}. Contributions of the counterterms with
the third order LECs are comparable, or even considerably lower than the
rest of the contributions to the third order.

Finally let us remark that both lessons learned from Table \ref{Tab thres1}
are confirmed by Table \ref{Tab thres2} --- the chiral expansion converges
to the experimental values, but the convergence seems to be rather slow, in a
sense that contributions to different orders are comparable. This fact seems
to show that despite of the relative success in describing elastic $\pi $N
scattering at threshold, the third order is definitely not the whole story.
A complete 1-loop calculation, which will include the fourth order of the
chiral expansion, is probably needed for sufficiently reliable description of
this process.

\begin{table} \centering
  \begin{tabular}{|c|c|c|c|} \hline
     &{\bf LECs}& {\bf rest}  & units\\ \hline
    $a^-_0 $      &  $\quad\   0.09 $ &  $\quad 0.03 $  & $GeV^{-1}$  \\ \hline
    $b^-_0 $     &  $\ -3.3 $      &  $\quad 0.5 $ & $GeV^{-3}$            \\ \hline
    $a^+_{1+} $      &  $ \quad 6.7$           &  $ -7.5 $        &   $GeV^{-3}$    \\ \hline
    $a^+_{1-} $  &  $ -13.4 $           &  $ -6.5 $         & $GeV^{-3}$      \\ \hline
    $a^-_{1+} $  &  $\ -6.7 $           &  $\quad 5.1 $   &  $GeV^{-3}$     \\ \hline
    $a^-_{1-} $  &  $\ -6.7 $            &  $ -4.5 $ & $GeV^{-3}$       \\ \hline
    $a^+_{2-} $  &  $\quad 1.9 $            &  $\ 24.7 $ &  $GeV^{-5}$       \\ \hline
    $a^-_{2-} $  &  $\quad 1.9 $            &  $\ 10.8 $ &  $GeV^{-5}$       \\ \hline
   \end{tabular}
  \caption{Comparison of contributions from counterterms with third order LECs and
contributions from loops + other counterterms (denoted together as "rest").
Comparison is made in the third order at the renormalization scale $\mu =1$GeV.
For the remaining 8 threshold parameters, not displayed in the Table,
contributions from counterterms with third-order LECs vanish.
\label{Tab comp}}
\end{table}

\newpage\

\section{Conclusions}

We have calculated the elastic $\pi $N scattering amplitude in the isospin
limit in the framework of HBCHPT, up to the third order. Since the chiral
expansion is supposed to work well near threshold, we have used the
extrapolated threshold parameters, like scattering lengths, volumes,
effective ranges, etc., to compare the results with data.

The elastic $\pi $N scattering amplitude and therefore also the threshold
parameters contain nine low energy constants (besides $F_\pi $ and $g_A$
from the lowest order Lagrangians). All these LECs were fixed from the
available pion-nucleon data --- the pion-nucleon $\sigma $-term,
Goldberger-Treiman discrepancy and the threshold parameters. Values of the
second order LECs are in a quite good agreement with their recent
determination in \cite{BKM96}. Values of the third order LECs were not
determined directly from $\pi $N data until now.

The third-order calculation brought a clear improvement in the description
of data, and this improvement was achieved with naturally small LECs. The
results, however, suggest importance of higher-order corrections, since the
contributions of the first three orders are frequently comparable and this
will probably be the case also in the fourth order.

\bigskip\ 

\noindent \TeXButton{Acknowledgements}{{\Large \bf Acknowledgements}}

\bigskip\ 

I owe much to G. Ecker, this work would be impossible without his continuous
help. I would also like to thank H. Leutwyler for useful comments to the
first version of the manuscript and R. Lietava and V. \v Cern\'y for helpful
discussions. And last but not least, a very useful correspondence with N.
Kaiser and U.-G. Mei\ss ner in the final stage of the work is acknowledged
with pleasure.

\newpage\ \

\appendix

\section{\label{rules}Feynman rules}

Propagators and vertices used in the calculation of elastic $\pi $%
N-scattering are summarized. Most of them can be found, e.g., in \cite{BKM95}%
, we present them here mainly for the sake of completeness and also because
of slightly different notation. Rules are given in the isospin limit $%
m_u=m_d.$ Nucleon momentum in final state $p^{\prime }$ is outgoing, all the
other momenta are ingoing. Momenta and isospins of pions are denoted as $%
\left( q_1,a\right) ,$ $\left( q_2,b\right) ,$ $\left( q_3,c\right) ,$ $%
\left( q_4,d\right) .$

\bigskip\

\noindent {\bf Propagators}

\medskip\ \

\noindent pion propagator
\begin{equation}
\label{pi prop}\frac{i\delta _{ab}}{k^2-M^2+i\varepsilon }
\end{equation}

\noindent nucleon propagator
\begin{equation}
\label{N prop}\frac i{v\cdot p+i\varepsilon }
\end{equation}

\bigskip\

\noindent ${\bf \pi }${\bf N vertices from }$\widehat{{\cal L}}_{\pi
N}^{(1)} $

\medskip\ \

\noindent one pion
\begin{equation}
\label{L1 1pi}-\frac{\stackrel{.}{g}_A}F\;S.q\;\tau _a
\end{equation}

\noindent two pions
\begin{equation}
\label{L1 2pi}\frac 1{4F^2}\;v.(q_1-q_2)\;\varepsilon _{abc}\tau _c
\end{equation}

\noindent three pions
\begin{equation}
\label{L1 3pi}-\frac{\stackrel{.}{g}_A}{2F^3}\left[ S.(q_1+q_2)\;\delta
_{ab}\tau _c+S.(q_1+q_3)\;\delta _{ac}\tau _b+S.(q_2+q_3)\;\delta _{bc}\tau
_a\right]
\end{equation}

\noindent four pions
\begin{equation}
\label{L1 4pi}
\begin{array}[t]{rl}
\frac i{2F^4} & v.(q_1+q_2+q_3+q_4)\;(\delta _{ab}\delta _{cd}+\delta
_{ac}\delta _{bd}+\delta _{ad}\delta _{bc})- \\
-\frac 1{\ 8F^4} & \left[ v.(q_2-q_1)\;\delta _{cd}\varepsilon _{abf}\tau
_f+v.(q_4-q_3)\;\delta _{ab}\varepsilon _{cdf}\tau _f+\right. \\
& \left. v.(q_3-q_1)\;\delta _{bd}\varepsilon _{acf}\tau
_f+v.(q_4-q_2)\;\delta _{ac}\varepsilon _{bdf}\tau _f+\right. \\
& \left. v.(q_4-q_1)\;\delta _{bc}\varepsilon _{adf}\tau
_f+v.(q_3-q_2)\;\delta _{ad}\varepsilon _{bcf}\tau _f\right]
\end{array}
\end{equation}

\newpage\ \

\noindent ${\bf \pi }${\bf N vertices from }$\widehat{{\cal L}}_{\pi
N}^{(2)} $

\medskip\

\noindent counterterm%
\begin{equation}
\label{L2 ct}i\left( \frac{p^2}{2m}+\frac{4M^2}ma_3\right)
\end{equation}

\noindent one pion
\begin{equation}
\label{L2 1pi}\frac{\stackrel{.}{g}_A}{2mF}\;v.q\;S.(p+p^{\prime })\;\tau _a
\end{equation}

\noindent two pions
\begin{equation}
\label{L2 2pi}
\begin{array}[t]{rl}
-\frac{4i}{mF^2} & \left( a_1\;q_1.q_2+a_2\;v.q_1\;v.q_2+a_3M^2\right)
\;\delta _{ab}+ \\
\frac 1{8mF^2} & \left[ (q_1-q_2)\cdot (p+p^{\prime })+i16a_5\varepsilon
^{\mu \nu \rho \sigma }q_{1\mu }q_{2\nu }v_\rho S_\sigma \right]
\;\varepsilon _{abc}\tau _c
\end{array}
\end{equation}

\bigskip\

\noindent ${\bf \pi }${\bf N vertices from }$\widehat{{\cal L}}_{\pi
N}^{(3)} $

\medskip\

\noindent one pion
\begin{equation}
\label{L3 1pi}
\begin{array}[t]{rl}
\frac{\stackrel{.}{g}_A}{8m^2F} & \left( q^2\;S\cdot q+2\;S\cdot p^{^{\prime
}}\;q\cdot p+2\;S\cdot p\;q\cdot p^{^{\prime }}\right) \;\tau _a+ \\
\frac{M^2}{8\pi ^2F^3} & \left( b_{19}-2b_{17}\right) \;S\cdot q\;\tau _a
\end{array}
\end{equation}

\noindent two pions
\begin{equation}
\label{L3 2pi}
\begin{array}[t]{cl}
\frac 1{8\pi ^2F^4} & \left\{ \left[ -ib_4(v\cdot q_2\;q_1\cdot (p+p^{\prime
})+v\cdot q_1\;q_2\cdot (p+p^{\prime }))+\right. \right. \\
& \left. \left. (b_{16}-b_{15})\varepsilon ^{\mu \nu \rho \sigma }q_{1\mu
}q_{2\nu }v_\rho S_\sigma \;v\cdot (q_1-q_2)+\right. \right. \\
& \left. \left.
\frac{\stackrel{.}{g}_A^2\pi ^2F^2}{m^2}(v\cdot q_1\;q_{2\mu }+v\cdot
q_2\;q_{1\mu })\varepsilon ^{\mu \nu \rho \sigma }(p-p^{^{\prime }})_\nu
v_\rho S_\sigma \right] \;\delta _{ab}+\right. \\  & \left. \left[
-(b_1+b_2)\;q_1\cdot q_2\;v\cdot \left( q_1-q_2\right) -\right. \right. \\
& \left. \left. b_3\;v\cdot q_1\;v\cdot q_2\;v\cdot \left( q_1-q_2\right)
+2b_6M^2\;v\cdot \left( q_1-q_2\right) +\right. \right. \\
& \left. \left. \frac{\pi ^2F^2}{m^2}(8a_5+3\stackrel{.}{g}%
_A^2-1)\varepsilon ^{\mu \nu \rho \sigma }q_{1\mu }q_{2\nu }(p+p^{^{\prime
}})_\rho S_\sigma \right] \;\varepsilon _{abc}\tau _c\right\}
\end{array}
\end{equation}

\bigskip\

\noindent {\bf 4}$\pi ${\bf -vertex from }${\cal L}_{\pi \pi }^{(2)}$

\begin{equation}
\label{L2 4pi}
\begin{array}[t]{rl}
\frac i{F^2}\;\{\!\! & [(q_1+q_2)^2-M^2]\delta _{ab}\delta _{cd}+ \\  
& [(q_1+q_3)^2-M^2]\delta _{ac}\delta _{bd}+ \\
& [(q_1+q_4)^2-M^2]\delta _{ad}\delta _{bc}\quad \}
\end{array}
\end{equation}

\bigskip\ 

\noindent {\bf pion counterterm vertex from }${\cal L}_{\pi \pi }^{(4)}$

\begin{equation}
\label{L4}2i\frac{M^2}{F^2}\left[ (q^2-M^2)l_4-M^2l_3\right] \delta _{ab} 
\end{equation}

\section{\label{loops}Loop integrals}

\noindent {\bf Definitions}
\begin{equation}
\label{defs}
\begin{array}{c}
\begin{array}{rcl}
\Delta & = & -\frac 1i\int
\frac{d^Dk}{\left( 2\pi \right) ^D}\frac 1{k^2-M^2} \\  &  &  \\
I_0(Q^2) & = & \frac 1i\int
\frac{d^Dk}{\left( 2\pi \right) ^D}\frac 1{\left( k^2-M^2\right) \left(
\left( k+Q\right) ^2-M^2\right) } \\  &  &  \\
Q_\mu I_1(Q^2) & = & \frac 1i\int
\frac{d^Dk}{\left( 2\pi \right) ^D}\frac{k_\mu }{\left( k^2-M^2\right)
\left( \left( k+Q\right) ^2-M^2\right) } \\  &  &  \\
g_{\mu \nu }I_2(Q^2)+Q_\mu Q_\nu I_3(Q^2) & = & \frac 1i\int
\frac{d^Dk}{\left( 2\pi \right) ^D}\frac{k_\mu k_\nu }{\left( k^2-M^2\right)
\left( \left( k+Q\right) ^2-M^2\right) } \\  &  &  \\
J_0\left( \omega \right) & = & \frac 1i\int
\frac{d^Dk}{\left( 2\pi \right) ^D}\frac 1{\left( k^2-M^2\right) \left(
\omega -v\cdot k\right) } \\  &  &  \\
v_\mu J_1\left( \omega \right) & = & \frac 1i\int
\frac{d^Dk}{\left( 2\pi \right) ^D}\frac{k_\mu }{\left( k^2-M^2\right)
\left( \omega -v\cdot k\right) } \\  &  &  \\
g_{\mu \nu }J_2\left( \omega \right) +v_\mu v_\nu J_3\left( \omega \right) &
= & \frac 1i\int
\frac{d^Dk}{\left( 2\pi \right) ^D}\frac{k_\mu k_\nu }{\left( k^2-M^2\right)
\left( \omega -v\cdot k\right) } \\  &  &  \\
K_0(\omega ,Q^2) & = & \frac 1i\int
\frac{d^Dk}{\left( 2\pi \right) ^D}\frac 1{\left( k^2-M^2\right) \left(
(k+Q)^2-M^2\right) \left( \omega -v\cdot k\right) } \\  &  &  \\
Q_\mu K_1(\omega ,Q^2)+v_\mu K_1^{^{\prime }}(\omega ,Q^2) & = & \frac
1i\int
\frac{d^Dk}{\left( 2\pi \right) ^D}\frac{k_\mu }{\left( k^2-M^2\right)
\left( (k+Q)^2-M^2\right) \left( \omega -v\cdot k\right) } \\  &  &
\end{array}
\\
\begin{array}{rcl}
g_{\mu \nu }K_2+Q_\mu Q_\nu K_3+ &  &  \\
&  &  \\
(v_\mu Q_\nu +Q_\mu v_\nu )K_3^{^{\prime }}+v_\mu v_\nu K_3^{^{\prime \prime
}} & = & \frac 1i\int \frac{d^Dk}{\left( 2\pi \right) ^D}\frac{k_\mu k_\nu }{%
\left( k^2-M^2\right) \left( (k+Q)^2-M^2\right) \left( \omega -v\cdot
k\right) }
\end{array}
\end{array}
\end{equation}

$$
\begin{array}{rcl}
(Q_\lambda g_{\mu \nu }+Q_\mu g_{\nu \lambda }+Q_\nu g_{\lambda \mu })K_4+ &
&  \\
&  &  \\
(v_\lambda g_{\mu \nu }+v_\mu g_{\nu \lambda }+v_\nu g_{\lambda \mu
})K_4^{\prime }+ &  &  \\
&  &  \\
(v_\lambda Q_\mu Q_\nu +v_\mu Q_\nu Q_\lambda +v_\nu Q_\lambda Q_\mu )K_5+ &
&  \\
&  &  \\
(Q_\lambda v_\mu v_\nu +Q_\mu v_\nu v_\lambda +Q_\nu v_\lambda v_\mu
)K_5^{^{\prime }}+ &  &  \\
&  &  \\
Q_\lambda Q_\mu Q_\nu K_6+v_\lambda v_\mu v_\nu K_6^{^{\prime }} & = & \frac
1i\int \frac{d^Dk}{\left( 2\pi \right) ^D}\frac{k_\lambda k_\mu k_\nu }{%
\left( k^2-M^2\right) \left( (k+Q)^2-M^2\right) \left( \omega -v\cdot
k\right) }
\end{array}
$$
\

\pagebreak

\noindent {\bf Results}\footnote{\noindent Divergent integrals are
calculated using dimensional regularization, with the terms in denominators
of integrals understood with the $i\varepsilon $ prescription from appendix
A (i.e. $\frac 1{k^2-M^2},$ $\frac 1{v\cdot k-\omega }$ etc. in definitions
are just abbreviations for $\frac 1{k^2-M^2+i\varepsilon },$ $\frac 1{v\cdot
k-\omega +i\varepsilon }$ etc). Results are given in the limit $D\rightarrow
4,$ i.e. in the final formulae the dimension of space-time $D=4-2\varepsilon
$ is expanded in powers of $\varepsilon ,$ then one uses $\varepsilon
L=-\frac 1{32\pi ^2}+{\cal O}(\varepsilon )$ and afterwards $\varepsilon $
is sent to zero.}

\bigskip 

$$
L(\mu )=\frac{\mu ^{D-4}}{16\pi ^2}\left\{ \frac 1{D-4}-\frac 12\left[ \ln
4\pi +1+\Gamma ^{\prime }(1)\right] \right\} 
$$

\bigskip\
\begin{equation}
\label{delta}\Delta =2M^2\left( L(\mu )+\frac 1{32\pi ^2}\ln \frac{M^2}{\mu
^2}\right) 
\end{equation}

\bigskip\ 
\begin{equation}
\label{I0}
\begin{array}[t]{cll}
I_0(Q^2)= & -2L(\mu )+\frac 1{16\pi ^2}\left( 1-\ln \frac{M^2}{\mu ^2}-r\ln
\left| \frac{1+r}{1-r}\right| \right) & \left[ Q^2<0\right] \\  
&  &  \\
& -2L(\mu )+\frac 1{16\pi ^2}\left( 1-\ln \frac{M^2}{\mu ^2}-2r\arctan \frac
1r\right) & \left[ 0<Q^2<4M^2\right] \\
&  &  \\  
& -2L(\mu )+\frac 1{16\pi ^2}\left( 1-\ln \frac{M^2}{\mu ^2}-r\ln \left|
\frac{1+r}{1-r}\right| +i\pi r\right) & \left[ Q^2>4M^2\right] \\  
&  &  \\  
& \text{where\qquad }r=\sqrt{\left| 1-\frac{4M^2}{Q^2}\right| } &
\end{array}
\end{equation}

\bigskip\ 
\begin{equation}
\label{J0}
\begin{array}[t]{cll}
J_0\left( \omega \right) = & -4\omega L(\mu )+\frac \omega {8\pi ^2}\left(
1-\ln \frac{M^2}{\mu ^2}\right) +\frac{\sqrt{\omega ^2-M^2}}{4\pi ^2}\;\text{%
arccosh\ }\frac{-\omega }M & \left[ \omega <-M\right] \\  
&  &  \\  
& -4\omega L(\mu )+\frac \omega {8\pi ^2}\left( 1-\ln \frac{M^2}{\mu ^2}%
\right) -\frac{\sqrt{M^2-\omega ^2}}{4\pi ^2}\arccos \frac{-\omega }M\
\qquad & \left[ \omega ^2<M^2\right] \\  
&  &  \\
& -4\omega L(\mu )+\frac \omega {8\pi ^2}\left( 1-\ln \frac{M^2}{\mu ^2}%
\right) -\frac{\sqrt{\omega ^2-M^2}}{4\pi ^2}\left( \text{arccosh\ }\frac
\omega M-i\pi \right) & \left[ \omega >M\right] 
\end{array}
\end{equation}

\bigskip\ 
\begin{equation}
\label{K0}
\begin{array}[t]{cll}
K_0(0,Q^2)= & -\frac 1{8\pi \sqrt{-Q^2}}\arctan \frac{\sqrt{-Q^2}}{2M} &
\left[ v\cdot Q=0 
\text{\quad and\quad }Q^2<0\right] \\  &  &  \\  
& \frac 1{16\pi \sqrt{Q^2}}\ln \frac{2M-\sqrt{Q^2}}{2M+\sqrt{Q^2}} & \left[
v\cdot Q=0 
\text{\quad and\quad }0<Q^2<4M^2\right] \\  &  &  \\  
& \frac 1{16\pi \sqrt{Q^2}}\left( \ln \frac{\sqrt{Q^2}-2M}{\sqrt{Q^2}+2M}%
+i\pi \right) & \left[ v\cdot Q=0\text{\quad and\quad }Q^2>4M^2\right]
\end{array}
\end{equation}

\pagebreak
\begin{equation}
\label{I1-3}
\begin{array}{lcl}
I_1(Q^2) & = & -\frac 12I_0(Q^2) \\
&  &  \\ 
I_2(Q^2) & = & \frac 13\left( \left( M^2- 
\frac{Q^2}4\right) I_0(Q^2)-\frac 12\Delta +\frac 1{16\pi ^2}\left( M^2-%
\frac{Q^2}6\right) \right) \qquad \\  &  &  \\ 
I_3(Q^2) & = & \frac 13\left( \left( 1-\frac{M^2}{Q^2}\right) I_0(Q^2)-\frac
1{2Q^2}\Delta -\frac 1{16\pi ^2}\left( \frac{M^2}{Q^2}-\frac 16\right)
\right) 
\end{array}
\end{equation}

\medskip\ \ 

\begin{equation}
\label{J1-3}
\begin{array}{lcl}
J_1\left( \omega \right) & = & \omega J_0\left( \omega \right) +\Delta \\
&  &  \\ 
J_2\left( \omega \right) \; & = & \frac 13\left( \left( M^2-\omega ^2\right)
J_0\left( \omega \right) -\omega \Delta +\frac 1{8\pi ^2}\left( \omega
M^2-\frac 23\omega ^3\right) \right) \qquad \\
&  &  \\ 
J_3\left( \omega \right) & = & \omega J_1\left( \omega \right) -J_2\left(
\omega \right) 
\end{array}
\end{equation}

\bigskip\ \ 
\begin{equation}
\label{K1-6}
\begin{array}{c}
\begin{array}{lcl}
K_1(0,Q^2) & = & -\frac 12K_0(0,Q^2) \\
&  &  \\ 
K_1^{^{\prime }}(0,Q^2) & = & -I_0(Q^2) \\
&  &  \\ 
K_2(0,Q^2) & = & \frac 12\left( M^2-\frac 14Q^2\right) K_0(0,Q^2)+\frac
14J_0(0) \\  
&  &  \\ 
K_3(0,Q^2) & = & \left( \frac 38-
\frac{M^2}{2Q^2}\right) K_0(0,Q^2)+\frac 1{4Q^2}J_0(0) \\  &  &  \\
K_3^{^{\prime }}(0,Q^2) & = & -I_1(Q^2) \\  
&  &  \\ 
K_3^{^{\prime \prime }}(0,Q^2) & = & -K_2(0,Q^2) \\  
&  &  \\ 
K_4(0,Q^2) & = & -\frac 14J_0(0)+\frac 12M^2K_1(0,Q^2)+\frac
14K_2(0,Q^2)+\frac 14Q^2K_3(0,Q^2) \\  
&  &  \\ 
K_4^{^{\prime }}(0,Q^2) & = & \frac 13\left( 
\frac{Q^2}4-M^2\right) I_0(Q^2)+\frac 16\Delta +\frac 1{48\pi ^2}\left( 
\frac{Q^2}6-M^2\right) \\  &  &  \\ 
K_5(0,Q^2) & = & -\frac 14I_0(Q^2)+\frac 1{2Q^2}\Delta -\frac
1{Q^2}K_4^{^{\prime }}(0,Q^2) \\
&  &  \\ 
K_5^{^{\prime }}(0,Q^2) & = & -K_4(0,Q^2)
\end{array}
\\
\begin{array}{ccl}
&  &  \\ 
K_6(0,Q^2)\; & = & -\frac 1{Q^2}J_0(0)+
\frac{M^2}{Q^2}K_1(0,Q^2)-\frac 5{Q^2}K_4(0,Q^2)\qquad \qquad \qquad \quad
\\  &  &  \\ 
K_6^{^{\prime }}(0,Q^2)\; & = & -I_2(Q^2)-3K_4^{^{\prime }}(0,Q^2) 
\end{array}
\end{array}
\end{equation}

\smallskip\ 

where all the{\bf \ }${\bf K}_i{\bf (0,Q}^2{\bf )}${\bf \ }are given only
{\bf for }${\bf v\cdot Q=0}$

\section{\label{diagrams}Feynman diagrams for $\pi $N scattering}

Amplitudes corresponding to Feynman diagrams in Fig.1 -- 4 are presented.
On-shell conditions
\begin{equation}
\label{on-shell}
\begin{array}{c}
q^2=q^{\prime 2}=M_\pi ^2 \\
v\cdot q-v\cdot q^{\prime }={\cal O}(p^2)
\end{array}
\end{equation}
are used to simplify expressions. Amplitudes are given in terms of loop
integrals defined in Appendix \ref{loops} and the following notation is used
\begin{equation}
\label{Jpm}J_n^{\pm }(w)=J_n(w)\pm J_n(-w) \quad .
\end{equation}

\medskip\ 

\noindent {\bf first order}

\medskip\

\noindent Fig. 1a 
\begin{equation}
\label{1a}
\begin{array}[t]{ccl}
\alpha ^{+} & = & \frac{\stackrel{.}{g}_A^2}{4F^2}\left( \frac 1w-\frac
1{w^{\prime }}\right) (q\cdot q^{\prime }-ww^{\prime }) \\ \beta ^{+} & = & 
\frac{\stackrel{.}{g}_A^2}{2F^2}\left( \frac 1w+\frac 1{w^{\prime }}\right)
\\ \alpha ^{-} & = & \frac{\stackrel{.}{g}_A^2}{4F^2}\left( \frac 1w+\frac
1{w^{\prime }}\right) (q\cdot q^{\prime }-ww^{\prime }) \\ \beta ^{-} & = & 
\frac{\stackrel{.}{g}_A^2}{2F^2}\left( \frac 1w-\frac 1{w^{\prime }}\right) 
\end{array}
\end{equation}

\smallskip 

\noindent Fig. 1b 
\begin{equation}
\label{1b}
\begin{array}[t]{ccl}
\alpha ^{+} & = & 0 \\ 
\beta ^{+} & = & 0 \\ 
\alpha ^{-} & = & \frac 1{4F^2}(w+w^{\prime }) \\ 
\beta ^{-} & = & 0 
\end{array}
\end{equation}

\medskip\ 

\noindent {\bf second order}

\medskip\ 

\noindent Fig. 2a 
\begin{equation}
\label{2a}
\begin{array}[t]{ccl}
\alpha ^{+} & = & - 
\frac{\stackrel{.}{g}_A^2}{4mF^2}\left( q\cdot q^{\prime }-ww^{\prime
}\right) \\ \beta ^{+} & = & 0 \\
\alpha ^{-} & = & 0 \\ 
\beta ^{-} & = & -\frac{\stackrel{.}{g}_A^2}{2mF^2} 
\end{array}
\end{equation}

\smallskip

\noindent Fig. 2b 
\begin{equation}
\label{2b}
\begin{array}[t]{ccl}
\alpha ^{+} & = & \frac{\stackrel{.}{g}_A^2}{8mF^2}\left[ 4ww^{\prime
}-w^2-w^{\prime 2}+(q\cdot q^{\prime }-2M_\pi ^2)\left( \frac{w^{^{\prime }}}%
w+\frac w{w^{^{\prime }}}\right) \right] \\ \beta ^{+} & = & \frac{\stackrel{%
.}{g}_A^2}{4mF^2}\left( \frac{w^{\prime }}w-\frac w{w^{\prime }}\right) \\ 
\alpha ^{-} & = & \frac{\stackrel{.}{g}_A^2}{8mF^2}\left[ w^2-w^{\prime
2}+(q\cdot q^{\prime }-2M_\pi ^2)\left( \frac{w^{^{\prime }}}w-\frac
w{w^{^{\prime }}}\right) \right] \\ \beta ^{-} & = & \frac{\stackrel{.}{g}%
_A^2}{4mF^2}\left( \frac{w^{\prime }}w+\frac w{w^{\prime }}\right)
\end{array}
\end{equation}

\smallskip\ 

\noindent Fig. 2c 
\begin{equation}
\label{2 c}
\begin{array}[t]{ccl}
\alpha ^{+} & = & - 
\frac{\stackrel{.}{g}_A^2M^2}{8mF^2}\left( \frac 1{w^2}+\frac 1{w^{\prime
2}}\right) (q\cdot q^{\prime }-ww^{\prime }) \\ \beta ^{+} & = & - 
\frac{\stackrel{.}{g}_A^2M^2}{4mF^2}\left( \frac 1{w^2}-\frac 1{w^{\prime
2}}\right) \\ \alpha ^{-} & = & - 
\frac{\stackrel{.}{g}_A^2M^2}{8mF^2}\left( \frac 1{w^2}-\frac 1{w^{\prime
2}}\right) (q\cdot q^{\prime }-ww^{\prime }) \\ \beta ^{-} & = & -\frac{%
\stackrel{.}{g}_A^2M^2}{4mF^2}\left( \frac 1{w^2}+\frac 1{w^{\prime
2}}\right) 
\end{array}
\end{equation}

\smallskip 

\noindent Fig. 2d 
\begin{equation}
\label{2d}
\begin{array}[t]{ccl}
\alpha ^{+} & = & \frac 4{mF^2}\left( a_1q.q^{\prime }+a_2ww^{\prime
}-a_3M^2\right)  \\ 
\beta ^{+} & = & 0 \\ 
\alpha ^{-} & = & 0 \\ 
\beta ^{-} & = & -\frac 2{mF^2}a_5
\end{array}
\end{equation}

\medskip\ 

\noindent {\bf third order}

\bigskip\

\medskip

\noindent Fig. 3a+3b 
\begin{equation}
\label{3ab}
\begin{array}[t]{ccl}
\alpha ^{+} & = & 0 \\ 
\beta ^{+} & = & \frac{\stackrel{.}{g}_A^2}{4m^2F^2}\frac 1w\left[ \frac{%
M^2m^2}{\pi ^2F^2\stackrel{.}{g}_A}(2b_{17}-b_{19})-M_\pi ^2+q\cdot
q^{^{\prime }}\right] \\ \alpha ^{-} & = & \frac{\stackrel{.}{g}_A^2}{8m^2F^2%
}\frac 1w\left[ \frac{M^2m^2}{\pi ^2F^2\stackrel{.}{g_A}}(2b_{17}-b_{19})-M_%
\pi ^2\right] \left( q\cdot q^{\prime }-w^2\right) \\  & + & \frac{\stackrel{%
.}{g}_A^2}{8m^2F^2}\frac 1w\;\left( q\cdot q^{^{\prime }}-M_\pi
^2+w^2\;\right) \left( q\cdot q^{^{\prime }}-M_\pi ^2\right) \\ \beta ^{-} & 
= & 0 
\end{array}
\end{equation}

\smallskip

\noindent Fig. 3c 
\begin{equation}
\label{3c}
\begin{array}[t]{ccl}
\alpha ^{+} & = & 0 \\ 
\beta ^{+} & = & -
\frac{\stackrel{.}{g}_A^2}{4m^2F^2}w \\ \alpha ^{-} & = & - 
\frac{\stackrel{.}{g}_A^2}{8m^2F^2}w\left( q\cdot q^{^{\prime }}+w^2-2M_\pi
^2\right) \\ \beta ^{-} & = & 0 
\end{array}
\end{equation}

\smallskip\ 

\noindent Fig. 3d

\begin{equation}
\label{3d}
\begin{array}[t]{ccl}
\alpha ^{+} & = & 0 \\
\beta ^{+} & = & \frac 1{4\pi ^2F^4}w\left[ b_{16}-b_{15}+ 
\frac{\stackrel{.}{g}_A^2\pi ^2F^2}{m^2}\right] \\ \alpha ^{-} & = & \frac
1{4\pi ^2F^4}w\left[ (b_1+b_2)\;q\cdot q^{\prime }+b_3w^2+2b_6M^2\right] \\ 
\beta ^{-} & = & 0 
\end{array}
\end{equation}

\noindent Fig. 3e+3f\TeXButton{nopagebreak}{\nopagebreak}

\begin{equation}
\label{3ef}
\begin{array}[t]{ccl}
\alpha ^{+} & = & 0 \\ 
\beta ^{+} & = & 0 \\ 
\alpha ^{-} & = & \frac{\stackrel{.}{g}_A^2M^2}{4m^2F^2}\frac 1w\left(
M^2-w^2\right) \\ \beta ^{-} & = & 0 
\end{array}
\end{equation}

\smallskip\

\noindent Fig. 3g

\begin{equation}
\label{3g}
\begin{array}[t]{ccl}
\alpha ^{+} & = & \frac{3g_A^4M^3}{64\pi F^4}\frac 1{w^2}\left( q\cdot
q^{\prime }-w^2\right) \\ \beta ^{+} & = & \frac{\stackrel{.}{g}_A^2M^2}{%
4m^2F^2}\frac 1w\left( \frac{M^2}{w^2}+16a_3\right) \\ \alpha ^{-} & = & 
\frac{\stackrel{.}{g}_A^2M^2}{8m^2F^2}\frac 1w\left( \frac{M^2}{w^2}%
+16a_3\right) \left( q\cdot q^{\prime }-w^2\right) \\ \beta ^{-} & = & \frac{%
3g_A^4M^3}{32\pi F^4}\frac 1{w^2} 
\end{array}
\end{equation}

\smallskip\ \ 

\noindent Fig. 4a, 4b%
$$
0
$$

\smallskip\ \

\noindent Fig. 4c+4d 
\begin{equation}
\label{4cd}
\begin{array}[t]{ccl}
\alpha ^{+} & = & - 
\frac{\stackrel{.}{g}_A^4}{8F^4}\frac 1{w^2}\left( J_2^{+}(w)-2J_2(0)\right)
\left( q\cdot q^{^{\prime }}-w^2\right) \\ \beta ^{+} & = & - 
\frac{\stackrel{.}{g}_A^4}{4F^4}\frac 1{w^2}\left( J_2^{-}(w)+\frac 1{6\pi
^2}w^3-\frac 1{4\pi ^2}M^2w\right) \\ \alpha ^{-} & = & - 
\frac{\stackrel{.}{g}_A^4}{8F^4}\frac 1{w^2}\left( J_2^{-}(w)+\frac 1{6\pi
^2}w^3-\frac 1{4\pi ^2}M^2w\right) \left( q\cdot q^{^{\prime }}-w^2\right)
\\ \beta ^{-} & = & -\frac{\stackrel{.}{g}_A^4}{4F^4}\frac 1{w^2}\left(
J_2^{+}(w)-2J_2(0)\right) 
\end{array}
\end{equation}

\smallskip 

\noindent Fig. 4e
\begin{equation}
\label{4e}
\begin{array}[t]{ccl}
\alpha ^{+} & = & \frac{3\stackrel{.}{g}_A^4}{16F^4}\frac 1{w^2}\left(
M^2-w^2\right) J_0^{+}(w)\left( q\cdot q^{^{\prime }}-w^2\right) \\ \beta
^{+} & = & \frac{3\stackrel{.}{g}_A^4}{8F^4}\frac 1{w^2}\left( \left(
M^2-w^2\right) J_0^{-}(w)-2w\Delta \right) \\ \alpha ^{-} & = & \frac{3%
\stackrel{.}{g}_A^4}{16F^4}\frac 1{w^2}\left( \left( M^2-w^2\right)
J_0^{-}(w)-2w\Delta \right) \left( q\cdot q^{^{\prime }}-w^2\right) \\ \beta
^{-} & = & \frac{3\stackrel{.}{g}_A^4}{8F^4}\frac 1{w^2}\left(
M^2-w^2\right) J_0^{+}(w) 
\end{array}
\end{equation}

\smallskip 

\noindent Fig. 4f
\begin{equation}
\label{4f}
\begin{array}[t]{ccl}
\alpha ^{+} & = & \frac 1{8F^4}\left( w^2J_0^{+}(w)+3wJ_1^{-}(w)\right) \\ 
\beta ^{+} & = & 0 \\ 
\alpha ^{-} & = & \frac 1{16F^4}\left( w^2J_0^{-}(w)+3wJ_1^{+}(w)\right) \\
\beta ^{-} & = & 0 
\end{array}
\end{equation}

\smallskip 

\noindent Fig. 4g, 4h%
$$
0 
$$

\medskip\ 

\noindent Fig. 4i
\begin{equation}
\label{4i}
\begin{array}[t]{ccl}
\alpha ^{+} & = & \frac{9\stackrel{.}{g}_A^4}{16F^4}\frac 1{w^2}\left(
J_2^{+}(w)-2J_2(0)\right) \left( q\cdot q^{^{\prime }}-w^2\right) \\ \beta
^{+} & = & \frac{3\stackrel{.}{g}_A^4}{8F^4}\frac 1{w^2}\left( -J_2^{-}(w)+2w%
\frac{\partial J_2(0)}{\partial \omega }+\frac 1{6\pi ^2}w^3\right) \\ 
\alpha ^{-} & = & \frac{3\stackrel{.}{g}_A^4}{16F^4}\frac 1{w^2}\left(
-J_2^{-}(w)+2w\frac{\partial J_2(0)}{\partial \omega }-\frac 1{18\pi
^2}w^3\right) \left( q\cdot q^{^{\prime }}-w^2\right) \\ \beta ^{-} & = & 
\frac{\stackrel{.}{g}_A^4}{8F^4}\frac 1{w^2}\left( J_2^{+}(w)-2J_2(0)\right) 
\end{array}
\end{equation}

\smallskip 

\noindent Fig. 4k 
\begin{equation}
\label{4k}
\begin{array}[t]{ccl}
\alpha ^{+} & = & 0 \\ 
\beta ^{+} & = & 0 \\ 
\alpha ^{-} & = & -\frac 1{F^4}wI_2(t)\qquad \qquad \; \\
\beta ^{-} & = & 0 
\end{array}
\end{equation}

\smallskip 

\noindent Fig. 4l 
\begin{equation}
\label{4l}
\begin{array}[t]{ccl}
\alpha ^{+} & = & \frac{\stackrel{.}{g}_A^2}{4F^4}\left[
(3t-M^2)tK_1(0,t)+(11t-3M^2)K_2(0,t)\right. \\  & + & \left.
(5t-M^2)tK_3(0,t)+10tK_4(0,t)+2t^2K_6(0,t)+6J_2(0)\right] \\ 
\beta ^{+} & = & 0 \\ 
\alpha ^{-} & = & \frac{\stackrel{.}{g}_A^2}{F^4}w\left[ tK_3^{^{\prime
}}(0,t)+tK_5(0,t)+(\frac t4-M^2)I_0(t)+\frac 12\Delta \right] \\ \beta ^{-}
& = & -\frac{\stackrel{.}{g}_A^2}{F^4}\;2K_2(0,t)
\end{array}
\end{equation}

\smallskip 

\noindent Fig. 4m
\begin{equation}
\label{4m}
\begin{array}[t]{ccl}
\alpha ^{+} & = & 0 \\ 
\beta ^{+} & = & 0 \\ 
\alpha ^{-} & = & \frac{\stackrel{.}{g}_A^2}{8F^4}w\left( 3\frac{\partial
J_2(0)}{\partial \omega }-\frac{M^2}{8\pi ^2}\right) \\ \beta ^{-} & = & 0 
\end{array}
\end{equation}

\smallskip 

\noindent Fig. 4n, 4o%
$$
0 
$$

\medskip\ 

\noindent Fig. 4p+4r
\begin{equation}
\label{4pr}
\begin{array}[t]{ccl}
\alpha ^{+} & = & 0 \\ 
\beta ^{+} & = & \frac{\stackrel{.}{g}_A^2}{F^4}\frac 1w\Delta \\ \alpha
^{-} & = & \frac{\stackrel{.}{g}_A^2}{2F^4}\frac 1w\Delta \left( q\cdot
q^{^{\prime }}-w^2\right) \\ \beta ^{-} & = & 0 
\end{array}
\end{equation}

\smallskip 

\noindent Fig. 4s%
$$
0 
$$

\medskip\ 

\noindent Fig. 4t+4u
\begin{equation}
\label{4tu}
\begin{array}[t]{ccl}
\alpha ^{+} & = & - 
\frac{3g_A^2}{2F^4}J_2(0)\; \\ \beta ^{+} & = & 0 \\ 
\alpha ^{-} & = & 0 \\ 
\beta ^{-} & = & 0 
\end{array}
\end{equation}

\smallskip 

\noindent Fig. 4v 
\begin{equation}
\label{4v}
\begin{array}[t]{ccl}
\alpha ^{+} & = & 0 \\
\beta ^{+} & = & 0 \\ 
\alpha ^{-} & = & \frac 5{8F^4}w\Delta  \\ 
\beta ^{-} & = & 0
\end{array}
\end{equation}

\smallskip\ \ 

\noindent external legs (wave function) renormalization

\begin{equation}
\label{ext}
\begin{array}[t]{ccl}
\alpha ^{+} & = & 0 \\ 
\beta ^{+} & = & \frac{\stackrel{.}{g}_A^2}{F^2}\frac 1w\left( \frac
12\delta Z_N(p)+\frac 12\delta Z_N(p^{\prime })+\delta Z_\pi \right) \; \\ 
\alpha ^{-} & = & \left[
\frac{\stackrel{.}{g}_A^2}{2F^2}\frac 1w\left( q\cdot q^{\prime }-w^2\right)
+\frac 1{2F^2}w\right] \left( \frac 12\delta Z_N(p)+\frac 12\delta
Z_N(p^{\prime })+\delta Z_\pi \right) \\ \beta ^{-} & = & 0
\end{array}
\end{equation}

$$
\begin{array}{ccl}
\delta Z_N(p) & = & \frac{4M_\pi ^2a_3}{m_N^2}-\frac{3g_A^2M_\pi ^2}{32\pi
^2F_\pi ^2}\left[ 1+48\pi ^2L(\mu )+3\ln \frac{M_\pi }\mu \right] \\ \delta
Z_N(p^{\prime }) & = & \frac{4M_\pi ^2a_3}{m_N^2}+\frac t{4m_N^2}-\frac{%
3g_A^2M_\pi ^2}{32\pi ^2F_\pi ^2}\left[ 1+48\pi ^2L(\mu )+3\ln \frac{M_\pi }%
\mu \right] \\ \delta Z_\pi & = & -\frac{M_\pi ^2}{8\pi ^2F_\pi ^2}\left[ 
\overline{l}_4+48\pi ^2L(\mu )+\ln \frac{M_\pi }\mu \right] 
\end{array}
$$

\newpage\

\section{\label{lengths}Scattering lengths, volumes, ...}

Threshold parameters for the first four partial waves are given in explicit
form, order by order. In the third order, the general formulae are displayed
only if they are reasonably short, otherwise only the numerical results are
given (in appropriate powers of GeV).

\medskip 

\noindent {\bf S-wave}\bigskip 

{\bf 1st order} 
\begin{equation}
\label{S1}
\begin{array}{rcl}
a_0^{+} & = & 0 \\  
&  &  \\ 
b_0^{+} & = & {0} \\  &  &  \\ 
a_0^{-} & = & {\frac{m_NM_\pi }{{8\pi F_\pi ^2\left( m_N+M_\pi \right) }}}
\\  &  &  \\
b_0^{-} & = & {\frac{2{{m_N^2}+3{M_\pi ^2}-4{g_A^2}M_\pi {\left( m_N+M_\pi
\right) }}}{32{\pi {F_\pi ^2}M_\pi {m_N}\left( m_N+M_\pi \right) }}} \\  &  
&  
\end{array}
\end{equation}

{\bf 2nd order} 
\begin{equation}
\label{S2}
\begin{array}{rcl}
a_0^{+} & = & {\frac{{\left( a_1+a_2-a_3\right) }M_\pi ^2}{{\pi {F}}_\pi ^2{%
\left( {m}_N+M_\pi \right) }}} \\  &  &  \\ 
b_0^{+} & = & {\frac{{a_1}\left( 4{{m}_N^2{{-2{m}_NM_\pi }+}M_\pi ^2}\right) 
{+a_2}\left( {4{m}_N^2+2{m}_NM_\pi {{+}5M_\pi ^2}}\right) {+a_3}\left( 2{{m}%
_NM_\pi -M_\pi ^2}\right) }{{4\pi {F}_\pi ^2{m}_N^2\left( {m}_N+M_\pi
\right) }}} \\  & + & \frac{{g_A^2}\left( {{2{{m}_N^2+}}3{m}_N}M{_\pi +3}%
M_\pi ^2\right) }{{16\pi {F}_\pi ^2{m}_N^2\left( {m}_N+M_\pi \right) }} \\  
&  &  \\ 
a_0^{-} & = & 0 \\  
&  &  \\
b_0^{-} & = & {\frac{{g_A^2M_\pi }}{{8\pi {F}_\pi ^2{{m}_N}\left( {m}%
_N+M_\pi \right) }}} \\  &  &  
\end{array}
\end{equation}

{\bf 3rd order}
\begin{equation}
\label{S3}
\begin{array}{rcl}
a_0^{+} & = & \frac{3g_A^2{{m}_N}M_\pi ^3}{256\pi ^2F_\pi ^4\left( m_N+M_\pi
\right) } \\  &  &  \\ 
b_0^{+} & = & \frac{g_A^2M_\pi \left( 154{{m}_N^2-18}m_NM_\pi +9M_\pi
^2\right) -g_A^4M_\pi ^2\left( 128m_N+64M_\pi \right) }{3072\pi ^2F_\pi ^4{{m%
}_N}\left( m_N+M_\pi \right) } \\  &  &  \\ 
a_0^{-} & = & \frac{M_\pi ^3a_3}{2\pi F_\pi ^2m_N\left( m_N+M_\pi \right) }+%
\frac{m_N^2M_\pi ^3\left[ \frac 14+\widetilde{b}_1+\widetilde{b}_2+%
\widetilde{b}_3+2\widetilde{b}_6\right] }{16\pi ^3F_\pi ^4m_N\left(
m_N+M_\pi \right) } \\  &  &  \\ 
b_0^{-} & = & -0.16+1.48a_3+5.15\left( 
\widetilde{b}_1+\widetilde{b}_2\right) +6.18\widetilde{b}_3+3.82\widetilde{b}%
_6+0.64b_{19} \\  &  &
\end{array}
\end{equation}

\newpage 

\noindent {\bf P-wave}\bigskip

{\bf 1st order} 
\begin{equation}
\label{P1}
\begin{array}{rcl}
a_{1+}^{+} & = & \frac{g_A^2}{{24\pi {F_\pi ^2}M_\pi }} \\  &  &  \\ 
a_{1-}^{+} & = & - 
{\frac{{g_A^2}}{{12\pi {F_\pi ^2}M_\pi }}} \\  &  &  \\ 
a_{1+}^{-} & = & {\frac{{{M_\pi ^{}}-2{g_A^2}\left( m_N+M_\pi \right) }}{{%
48\pi {F_\pi ^2}M_\pi \left( m_N+M_\pi \right) }}} \\  &  &  \\ 
a_{1-}^{-} & = & {\frac{2m_NM_\pi {-3{M_\pi ^2}-{4g_A^2}m_N{\left( m_N+M_\pi
\right) }}}{{96\pi {F_\pi ^2}M_\pi m_N\left( m_N+M_\pi \right) }}} \\  &  &  
\end{array}
\end{equation}

{\bf 2nd order} 
\begin{equation}
\label{P2}
\begin{array}{rcl}
a_{1+}^{+} & = & {\frac{-{16{{m}_N}}a_1+{16M_\pi }a_2{+{g_A^2}}\left( {{{m}%
_N+}3M_\pi }\right) }{{48\pi {F}_\pi ^2{{m}_N}\left( {m}_N+M_\pi \right) 
}}} \\  &  &  \\ 
a_{1-}^{+} & = & {\frac{-a_1\left( {16{{m}_N^2+{12}M_\pi ^2}}\right)
+a_2\left( {16{m}_NM_\pi -12{M_\pi ^2}}\right) +12{{{{a_3}M_\pi ^2}}+{{g_A^2m%
}_N}}\left( {{{m}_N+}3M_\pi }\right) }{{48\pi {F}_\pi ^2{{m}_N^2}\left( {%
m}_N+M_\pi \right) }}} \\  &  &  \\ 
a_{1+}^{-} & = & - 
\frac{{4}a_5\left( {{m}_N{+}M_\pi }\right) {+{g_A^2}}\left( {{{m}_N}}+3M_\pi
\right) }{{48\pi {F}_\pi ^2{{m}_N}\left( {m}_N+M_\pi \right) }} \\  &  &  \\ 
a_{1-}^{-} & = & {\frac{{4a_5\left( {m}_N+M_\pi \right) +{g_A^2}m}_N}{{24\pi 
{F}_\pi ^2{{m}_N}\left( {m}_N+M_\pi \right) }}} \\  &  &  
\end{array}
\end{equation}

{\bf 3rd order} 
\begin{equation}
\label{P3}
\begin{array}{rcl}
a_{1+}^{+} & = & -6.47+0.62\left( b_{16}-
\widetilde{b}_{15}\right) -0.78b_{19} \\  &  &  \\ 
a_{1-}^{+} & = & -8.60-1.24\left( b_{16}- 
\widetilde{b}_{15}\right) +1.56b_{19} \\  &  &  \\ 
a_{1+}^{-} & = & 2.61-1.08\left( 
\widetilde{b}_1+\widetilde{b}_2\right) +0.68b_{19} \\  &  &  \\ 
a_{1-}^{-} & = & -6.94-0.014a_3-1.10\left( 
\widetilde{b}_1+\widetilde{b}_2\right) -0.018\widetilde{b}_3-0.035\widetilde{%
b}_6+0.68b_{19} \\  &  &  
\end{array}
\end{equation}

\newpage 

\noindent {\bf D-wave}\bigskip

{\bf 1st order} 
\begin{equation}
\label{D1}
\begin{array}{rcl}
a_{2+}^{+} & = & -
{\frac{{g_A^2}}{{60\pi {F_\pi ^2M_\pi ^2}m_N}}} \\  &  &  \\ 
a_{2-}^{+} & = & {\frac{g_A^2{\left( 2m_N+5M_\pi \right) }}{{480\pi {F_\pi
^2M_\pi ^2}m_N^2}}} \\  &  &  \\ 
a_{2+}^{-} & = & \frac{{g_A^2}}{{60\pi {F_\pi ^2M_\pi ^2}m_N}} \\  &  &  \\ 
a_{2-}^{-} & = & - 
{\frac{{g_A^2\left( 4m_N^2-6m_NM_\pi -10M_\pi ^2\right) +}5M_\pi ^2}{96{0\pi 
{F_\pi ^2M_\pi ^2{m_N^2}}\left( m_N+M_\pi \right) }}} \\  &  &  
\end{array}
\end{equation}

{\bf 2nd order} 
\begin{equation}
\label{D2}
\begin{array}{rcl}
a_{2+}^{+} & = & -
{\frac{{{g_A^2}\left( {{m}_N}+2M_\pi \right) }}{{120\pi {F}_\pi ^2M_\pi {{m}%
_N^2}\left( {m}_N+M_\pi \right) }}} \\  &  &  \\ 
a_{2-}^{+} & = & \frac{{80a_1{m}_NM_\pi -80a_2M_\pi ^2-{g_A^2}}\left( {8{{m}%
_N^2+}21{m}_NM_\pi +15}M_\pi ^2\right) }{{960\pi {F}_\pi ^2M_\pi {m}_N^3}%
\left( {m}_N{+M_\pi }\right) } \\  &  &  \\
a_{2+}^{-} & = & {\frac{{g_A^2\left( {{m}_N}+2M_\pi \right) }}{{120\pi {F}%
_\pi ^2{\,M_\pi {m}_N^2}\left( {m}_N+M_\pi \right) }}} \\  &  &  \\ 
a_{2-}^{-} & = & - 
{\frac{20a_5{M_\pi \left( {m}_N+M_\pi \right) }+{g_A^2}\left( {12{\,{m}%
_N^2+9m}_NM_\pi -5}M_\pi ^2\right) }{{960\pi {F}_\pi ^2{\,}M_\pi {{m}_N^3}%
\left( {m}_N+M_\pi \right) }}} \\  &  &  
\end{array}
\end{equation}

{\bf 3rd order} 
\begin{equation}
\label{D3}
\begin{array}{rcl}
a_{2+}^{+} & = & \frac{193g_A^2m_N}{115200\pi ^2F_\pi ^4M_\pi \left(
m_N+M_\pi \right) } \\  &  &  \\
a_{2-}^{+} & = & 25.0+0.18\left( b_{16}- 
\widetilde{b}_{15}\right) -0.22b_{19} \\  &  &  \\ 
a_{2+}^{-} & = & \frac{\left[ 1+g_A^2\left( 7-5\pi \right) \right] m_N-5\pi
g_A^2M_\pi }{14400\pi ^3F_\pi ^4M_\pi \left( m_N+M_\pi \right) } \\  &  &  
\\
a_{2-}^{-} & = & 11.5+0.306\left( 
\widetilde{b}_1+\widetilde{b}_2\right) -0.193b_{19} \\  &  &  
\end{array}
\end{equation}

\newpage 

\noindent {\bf F-wave}\bigskip

{\bf 1st order} 
\begin{equation}
\label{F1}
\begin{array}{rcl}
a_{3+}^{+} & = & \frac{g_A^2}{140\pi F_\pi ^2M_\pi ^3m_N^2} \\  &  &  \\
a_{3-}^{+} & = & -
\frac{g_A^2}{840\pi F_\pi ^2M_\pi ^3m_N^2} \\  &  &  \\ 
a_{3+}^{-} & = & - 
\frac{g_A^2}{140\pi F_\pi ^2M_\pi ^3m_N^2} \\  &  &  \\ 
a_{3-}^{-} & = & \frac{g_A^2}{840\pi F_\pi ^2M_\pi ^3m_N^2} \\  &  &  
\end{array}
\end{equation}

{\bf 2nd order} 
\begin{equation}
\label{F2}
\begin{array}{rcl}
a_{3+}^{+} & = & \frac{g_A^2\left( 2m_N+3M_\pi \right) }{280\pi F_\pi
^2M_\pi ^2m_N^3\left( m_N+M_\pi \right) } \\  &  &  \\ 
a_{3-}^{+} & = & - 
\frac{g_A^2\left( 4m_N^2-15m_NM_\pi -14M_\pi ^2\right) }{3360\pi F_\pi
^2M_\pi ^2m_N^4\left( m_N+M_\pi \right) } \\  &  &  \\ 
a_{3+}^{-} & = & - 
\frac{g_A^2\left( 2m_N+3M_\pi \right) }{280\pi F_\pi ^2M_\pi ^2m_N^3\left(
m_N+M_\pi \right) } \\  &  &  \\
a_{3-}^{-} & = & \frac{g_A^2\left( 4m_N-M_\pi \right) }{3360\pi F_\pi
^2M_\pi ^2m_N^3\left( m_N+M_\pi \right) } \\  &  &  
\end{array}
\end{equation}

{\bf 3rd order}
\begin{equation}
\label{F3}
\begin{array}{rcl}
a_{3+}^{+} & = & \frac{73g_A^2m_N}{752640\pi ^2F_\pi ^4M_\pi ^3\left(
m_N+M_\pi \right) } \\  &  &  \\ 
a_{3-}^{+} & = & \frac{g_A^2\left( 2190m_N^2-9457M_\pi ^2\right) }{%
22579200\pi ^2F_\pi ^4M_\pi ^3m_N\left( m_N+M_\pi \right) } \\  &  &  \\ 
a_{3+}^{-} & = & \frac{m_N\left[ 2+g_A^2\left( 18-7\pi \right) \right] -7\pi
g_A^2M_\pi }{470400\pi ^3F_\pi ^4M_\pi ^3\left( m_N+M_\pi \right) } \\  &  &
\\ 
a_{3-}^{-} & = & 23.3 \\  
&  &
\end{array}
\end{equation}


\newpage


\begin{thebibliography}{99}
\bibitem{W66}  S. Weinberg, Phys. Rev. Lett. 17 (1966) 616

\bibitem{GL84}  J. Gasser and H. Leutwyler, Ann. Phys. (N.Y.) 158 (1984)
142; Nucl. Phys. B250 (1985) 465

\bibitem{GSS88}  J. Gasser, M.E. Sainio and A. \v Svarc, Nucl. Phys. B307
(1988) 779

\bibitem{W79}  S. Weinberg, Physica 96A(1979) 327

\bibitem{JM91}  E. Jenkins and A.V. Manohar, Phys. Lett. B255 (1991) 558

\bibitem{BKM95}  V. Bernard, N. Kaiser and U.-G. Mei\ss ner, Int. J. Mod.
Phys. E4 (1995) 193

\bibitem{BKM95b}  V. Bernard, N. Kaiser and U.-G. Mei\ss ner, Phys. Rev. C52
(1995) 2185

\bibitem{BM96}  B. Borasoy and U.-G. Mei\ss ner, Ann. Phys.(NY) 254 (1997) 192

\bibitem{BKM93}  V. Bernard, N. Kaiser and U.-G. Mei\ss ner, Phys. Lett.
B309 (1993) 421

\bibitem{BKM96}  V. Bernard, N. Kaiser and U.-G. Mei\ss ner, Nucl. Phys. A615
(1997) 483

\bibitem{EGPdR89}  G. Ecker, J. Gasser, A. Pich and E. de Rafael, Nucl.
Phys. B321 (1989) 311

\bibitem{MRR92}  T. Mannel, W. Roberts and Z. Ryzak, Nucl. Phys. B368 (1992)
315

\bibitem{EM96}  G. Ecker and M. Moj\v zi\v s, Phys. Lett. B365 (1996) 312

\bibitem{BKKM92}  V. Bernard, N. Kaiser, J. Kambor and U.-G. Mei\ss ner,
Nucl. Phys. B388 (1992) 315

\bibitem{FLMS97}  H. Fearing, R. Lewis, N. Mobed and S. Scherer, Muon
capture by a proton in heavy baryon chiral perturbation theory, hep-ph/9702394

\bibitem{EM97}  G. Ecker and M. Moj\v zi\v s, Wave function renormalization in 
heavy baryon chiral perturbation theory, hep-ph/9705216

\bibitem{E94}  G. Ecker, Phys. Lett. B336 (1994) 508

\bibitem{H83}  G. H\"ohler, in Landolt-B\"ornstein, vol. 9 b2, ed. H.
Schopper (Springer, Berlin, 1983)

\bibitem{H96}  G. H\"ohler and H.M.Staudenmaier, $\pi $N Newsletter 11 (1996)

\bibitem{K80}  R. Koch and E. Pietarinen, Nucl. Phys. A336 (1980) 331
\end{thebibliography}
\end{document}